\documentclass[a4paper,fleqn,usenatbib]{mnras}  

\usepackage{mathptmx}

\usepackage[T1]{fontenc}
\usepackage{ae,aecompl}
\usepackage{multirow}
\usepackage{pdflscape}
\usepackage{ulem}

\usepackage{graphicx}	
\usepackage{amsmath}	
\usepackage{amssymb}	

\newcommand{\tbf}{\textbf}
\newcommand{\angstrom}{\mbox{\normalfont\AA}}

\title[RM Lag Uncertainties]{On Reverberation Mapping Lag Uncertainties}

\author[Yu et al.]{
Zhefu Yu$^{1}$,
C.~S.~Kochanek$^{1,2}$,
B.~M.~Peterson$^{1,2,3}$,
Y.~Zu$^{4}$,
W.~N.~Brandt$^{5,6,7}$,
\newauthor
E.~M.~Cackett$^{8}$,
M.~M.~Fausnaugh$^{9}$,
I.~M.~McHardy$^{10}$
\\
$^{1}$Department of Astronomy, The Ohio State University, Columbus, Ohio 43210, USA\\
$^{2}$Center of Cosmology and Astro-Particle Physics, The Ohio State University, Columbus, Ohio, 43210, USA\\
$^{3}$Space Telescope Science Institute, 3700 San Martin Drive, Baltimore, MD 21218, USA\\
$^{4}$Department of Astronomy, School of Physics and Astronomy, Shanghai Jiao Tong University, Shanghai 200240, China\\
$^{5}$Department of Astronomy and Astrophysics, Eberly College of Science, The Pennsylvania State University, \\
525 Davey Laboratory, University Park, PA 16802, USA\\
$^{6}$Department of Physics, The Pennsylvania State University, 104 Davey Laboratory, University Park, PA 16802, USA\\
$^{7}$Institute for Gravitation and the Cosmos, The Pennsylvania State University, University Park, PA 16802, USA\\
$^{8}$Department of Physics and Astronomy, Wayne State University, 666 W. Hancock St, Detroit, MI 48201, USA\\
$^{9}$Kavli Institute for Space and Astrophysics Research, Massachusetts Institute of Technology, 77 Massachusetts Avenue, \\
Cambridge, MA 02139-4307, USA\\
$^{10}$University of Southampton, Highfield, Southampton, SO17 1BJ, UK
}

\date{Accepted 2019 December 5. Received 2019 December 4; in original form 2019 September 4}

\pubyear{2020}

\begin{document}
\label{firstpage}
\pagerange{\pageref{firstpage}--\pageref{lastpage}}
\maketitle

\begin{abstract}
We broadly explore the effects of systematic errors on reverberation mapping lag uncertainty estimates from {\tt JAVELIN} and the interpolated cross-correlation function (ICCF) method. We focus on simulated lightcurves from random realizations of the lightcurves of five intensively monitored AGNs. Both methods generally work well even in the presence of systematic errors, although {\tt JAVELIN} generally provides better error estimates. Poorly estimated lightcurve uncertainties have less effect on the ICCF method because, unlike {\tt JAVELIN}, it does not explicitly assume Gaussian statistics. Neither method is sensitive to changes in the stochastic process driving the continuum or the transfer function relating the line lightcurve to the continuum. The only systematic error we considered that causes significant problems is if the line lightcurve is not a smoothed and shifted version of the continuum lightcurve but instead contains some additional sources of variability. 
\end{abstract}




\section{Introduction} \label{sec:intro}
The masses of supermassive black holes (SMBHs) are critical to understanding active galactic nuclei (AGNs), their evolution and their effect on host galaxies. In nearby normal galaxies, direct SMBH mass measurements can be made using the kinematics of stars \citep[e.g.,][]{vanderMarel1994,Gebhardt2009} or gas \citep[e.g.,][]{Harms1994,Barth2016}. These techniques require both high spatial resolution to resolve the black hole's region of influence and that the accretion activity is low enough to allow observations of the stars and gas. This restricts these measurements to nearby, inactive or mildly active galaxies. In AGNs, the reverberation mapping (RM) technique provides an approach to measuring the black hole mass using variability. Without the need for spatial resolution, RM allows SMBH mass measurements in active galaxies at (in principle) any distance. 

RM follows the response of the broad line region (BLR) emission lines to the variations in the continuum emission from the accretion disk. We can express the relation between the emission line and the continuum variations using a ``transfer function'' \footnote{$\Psi(v,\tau)$ is also referred as the ``response function'', depending on whether it is weighted by emissivity or responsivity. It is not critical to distinguish between these terms for this paper, so we only use the term ``transfer function'' here.} $\Psi(v,\tau)$  for the response of the line emission with line-of-sight velocity $v$ after a time delay $\tau$ from a change in the continuum. Resolving the velocity dependence of $\Psi(v,\tau)$ requires high cadence and signal-to-noise data \citep[e.g.,][]{Blandford1982,Horne2004}, so most RM studies consider only a one-dimensional ``delay map'' $\Psi(\tau)$ for the overall response of the line. In a linear echo model, the emission-line lightcurve is 
\begin{equation}
L(t) = L_0 + \int \Psi(\tau) \, \Delta C(t-\tau)\,d\tau 
\label{eq:linelc}
\end{equation}
where $L_0$ is a constant that depends on the non-varying continuum level, $\Psi(\tau)$ is the transfer function and $\Delta C(t-\tau)$ is the varying component of the continuum. The mean (centroid) time lag
\begin{equation}
\langle \tau \rangle = \frac{\int_0^{\infty} t \Psi(t) \, dt}{\int_0^{\infty} \Psi(t) \, dt}
\label{eq:meanlag}
\end{equation}
is then related to the black hole mass by
\begin{equation}
M_{\rm BH} = \frac{f c \langle \tau \rangle \Delta v^2}{G}
\label{eq:bhmass}
\end{equation}
where $f$ is a dimensionless ``virial factor'' determined by the structure and kinematics of the BLR and $\Delta v$ is the width of the broad emission line.

In addition to RM studies of emission line lags, continuum RM studies measured the lags between different wavelengths of the continuum. The standard thin accretion disk model is hottest near the center and colder at larger radii \citep[e.g.][]{Shakura1973,Shields1978}. If the continuum variability is driven by variable irradiation from the central regions, variability at longer wavelength will lag that at shorter wavelength. The continuum lag therefore encodes the size of the accretion disk as a function of temperature. Continuum RM studies have yielded lag measurements from both intensively monitored nearby AGNs \citep[e.g.,][]{Shappee2014,Edelson2015,Fausnaugh2016,Cackett2018,McHardy2018} and more distant objects from large sky surveys \citep[e.g.,][]{Jiang2017,Mudd2018,Homayouni2018,Yu2018}. 

Various algorithms have been used to estimate lags, such as the interpolated cross-correlation function \citep[ICCF, e.g.,][]{Gaskell1987,Peterson1998,Peterson2004}, the discrete cross-correlation function \citep[DCF, e.g.,][]{Edelson1988}, regularized linear inversion \citep[e.g.,][]{Krolik1995,Skielboe2015}, the z-transformed cross-correlation function \citep[ZDCF, e.g.,][]{Alexander1997,Alexander2013}, the Fourier cross-spectrum \citep[mainly for X-ray RM, e.g.,][]{Zhang2002,Uttley2014,Epitropakis2016}, {\tt JAVELIN} \citep[e.g.,][]{Zu2011,Zu2013} and {\tt CREAM} \citep[e.g.,][]{Starkey2016}. There have been many comparisons of these methods \citep[e.g.,][]{Koen1994,Kovacevic2014,King2015,Li2019}. Here we focus on the effects of systematic errors for the two most commonly used algorithms, the ICCF method and JAVELIN.

The ICCF method linearly interpolates the lightcurves and calculates the CCF. Either the centroid $\tau_{\rm cent}$ or the peak $\tau_{\rm peak}$ of the CCF can be an estimate of the time lag. For the lag uncertainty, the algorithm randomly picks a subset of the epochs (with replacement) and/or randomizes the flux to create a number of independent realizations of the lightcurves. These realizations build up the cross-correlation centroid distribution (CCCD) and cross-correlation peak distribution (CCPD), and the widths of these distributions are used as the estimate of the lag uncertainty.  

{\tt JAVELIN} combines an approach originally introduced for gravitational lensing time delays \citep{Press1992a,Press1992b} with recent statistical models for quasar variability \citep[e.g.,][]{Kelly2009,Kozlowski2010,MacLeod2010,Zu2013}. It models the AGN variability using a damped random walk (DRW) with a covariance function 
\begin{equation}
S(\Delta t) = \sigma_{\mbox{\tiny DRW}}^2 \, {\rm exp}(-|\Delta t / \tau_{\mbox{\tiny DRW}}|)
\label{eq:cov_drw}
\end{equation}
where $\sigma_{\mbox{\tiny DRW}}$ and $\tau_{\mbox{\tiny DRW}}$ are the amplitude and characteristic time scale, respectively. The DRW is a ``red noise'' process at short time scales with a power spectral density (PSD) slope of $-$2. The PSD flattens on time scales much larger than $\tau_{\mbox{\tiny DRW}}$. {\tt JAVELIN} assumes that the line lightcurve is a shifted, smoothed and scaled version of the continuum lightcurve (i.e., Equation \ref{eq:linelc}), and fits for the time lag, the width of a top-hat transfer function and the scaling factor that best reproduces the lightcurves using a Markov chain Monte Carlo (MCMC) algorithm. It estimates the lag uncertainty as the width of the posterior probability density distribution.  

A number of studies have noted that ICCF and {\tt JAVELIN} tend to derive different uncertainties, generally in the sense that the ICCF error estimates are larger \citep[e.g.,][]{Fausnaugh2017,Grier2017,McHardy2018,Mudd2018,Czerny2019,Edelson2019}. This has driven a range of speculations as to both the origin of the difference and as to which estimates are more reliable. Some considerations are the effect of incorrect lightcurve error estimates, deviations of quasar variability from the DRW model and choices of the transfer function (a top-hat by default in {\tt JAVELIN}). 

Correct lag uncertainty estimates are critical to the RM method. For example, lag uncertainty estimates directly affect the estimates of the intrinsic scatter in the scaling relation between the BLR size and the continuum luminosity \citep[e.g.,][]{Kaspi2000,Bentz2013}, which is widely used in single-epoch black hole mass estimates. Correct continuum lag uncertainties are important in constraining the accretion models and understanding the apparent discrepancy between the thin disk model and some observations \citep[e.g.,][]{Shappee2014,Fausnaugh2016,Jiang2017}. Therefore, a systematic study of these issues for the lag uncertainty estimates is necessary.

In this paper, we use observationally constrained simulated lightcurves to probe the effect of a broad range of systematic errors on the {\tt JAVELIN} and ICCF methods. We focus on high-cadence lightcurves and do not consider sampling strategies or gaps due to weather,
diurnal cycle, etc., as these have been examined in detail in previous studies \citep[e.g.,][]{Horne2004,King2015,Shen2015,Yu2018,Li2019}. The structure of the paper is as follows. Section \ref{sec:method} describes the observations used to build the simulated lightcurves and the simulation methodology. In Section \ref{sec:anl} we discuss the {\tt JAVELIN} and ICCF results for all the different model configurations. We summarize our findings in Section \ref{sec:summary}.

\section{Methodology} \label{sec:method}
\begin{figure*}
\includegraphics[width=\linewidth]{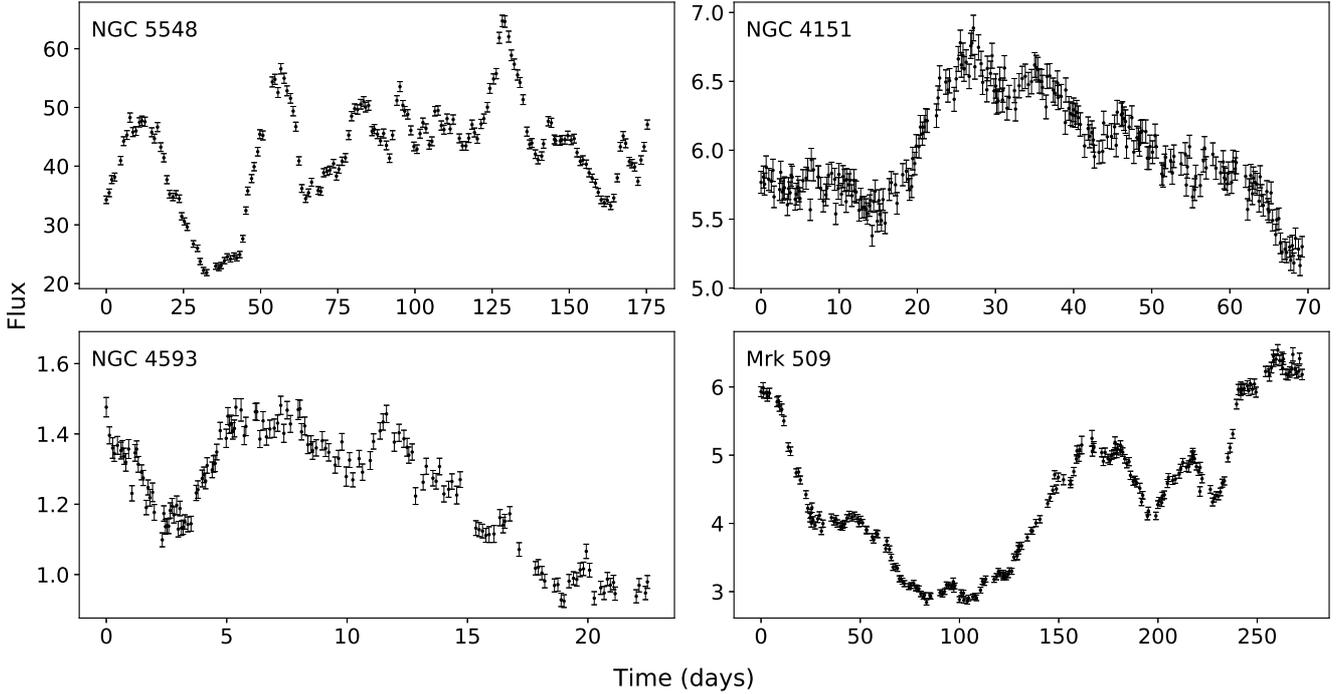}
\caption{The HST 1367\AA\ lightcurves of NGC 5548 (upper left) and the Swift UVW2-band lightcurves of NGC 4151 (upper right), NGC 4593 (bottom left) and Mrk 509 (bottom right) used in our experiments.}
\label{fig:obslc}
\end{figure*}

We base most of our simulations on the observed continuum lightcurves of four AGNs: NGC 5548, NGC 4151, NGC 4593 and Mrk 509. We show these observed lightcurves in Figure \ref{fig:obslc}. For NGC 5548, we adopt the 1367\AA\ lightcurve from the Hubble Space Telescope (HST) as part of the AGN Space Telescope and Optical RM (AGN STORM) Project \citep{DeRosa2015}. HST monitored NGC 5548 with the Cosmic Origins Spectrograph from February 17, 2014 to July 22, 2014. The lightcurve includes 171 epochs with a typical cadence of about one day. We use the Swift UVW2-band lightcurves for the other three AGNs. Swift monitored NGC 4151 in 2016 from February 20 to April 29, yielding a lightcurve that consists of 319 visits with nearly five visits per day \citep{Edelson2017}. The lightcurve of NGC 4593 contains 148 epochs with a cadence of about 96 minutes from July 13 to July 18 in 2016 and a cadence of about 192 minutes in the following 16.2 days \citep{McHardy2018}. The observations of Mrk 509 span from March 17 to December 15 in 2017 with 257 epochs separated by about one day \citep{Edelson2019}. We also carry out several tests using DRW lightcurves unconstrained by these observed lightcurves or using the Kepler lightcurve of Zw 229$-$15 \citep{Edelson2014} for the simulated continuum.      

We create simulated continuum lightcurves constrained by the observed lightcurves following the formalism of \citet{Zu2011} from \citet{Rybicki1992} based on theories of interpolation and prediction with a Gaussian process \citep[e.g.,][]{Lewis1971,Rao1973,OHagan1978}. Let vector $\tbf{y} = (y_1,\, y_2\, ... \, y_{N_{p}})$ represent the lightcurve with $N_{p}$ data points. The lightcurve $\tbf{y} = \tbf{s} + \tbf{n} + L\tbf{q}$ is a combination of the intrinsic signal $\tbf{s}$, the noise $\tbf{n}$ and a general trend $L\tbf{q}$. If the systematic trend is a constant background, $L$ is a $N_{p}\times1$ matrix with all elements equal to one and $\tbf{q}$ is the best-fit constant flux. While we only use a constant background, the linear parameters can also be used as a method to detrend the lightcurve as a polynomial ($a_0 + a_1 t + a_2 t^2$ etc.) or any other linear combination of functions. We define the signal covariance matrix $S=\langle \tbf{s}\tbf{s} \rangle$ and the noise covariance matrix $N=\langle \tbf{n}\tbf{n} \rangle$. The signal covariance matrix $S$ depends on the assumed stochastic process for the quasar variability. If the noise is uncorrelated, the noise covariance matrix $N$ is diagonal, and the diagonal elements are $N_{ii} = \sigma_{i}^2$, where $\sigma_{i}$ is the measurement error for the $i$th epoch.

\citet{Rybicki1992} showed that for a given observed lightcurve $\tbf{y}$, a signal covariance matrix $S$ specified by the assumed stochastic process and a noise covariance matrix $N$, the least-squares estimate of the mean of lightcurves consistent with the data is 
\begin{equation}
\hat{\tbf{s}} = SC^{-1}(\tbf{y} - L\hat{\tbf{q}})\
\label{eq:lc_bestfit}
\end{equation}
and the best-fit linear coefficients are
\begin{equation}
\hat{\tbf{q}} = C_q L^T C^{-1} \tbf{y}
\label{eq:q_bestfit}
\end{equation}
where $C = S + N$ and $C_q = (L^T C^{-1} L)^{-1}$. The dispersion of the lightcurves around the mean is 
\begin{equation}
\langle \Delta \tbf{s}^2 \rangle = S - S^T C_{\perp} S
\label{eq:std_bestfit}
\end{equation}
where $C_{\perp}^{-1} = C^{-1} - C^{-1} L C_q L^T C^{-1}$. Simulated lightcurves constrained by an observed lightcurve can also be constructed. The matrices $S$ and $N$ are given entries for the epochs both with and without data, while the noise is set to infinity (i.e. the corresponding entry in $N^{-1}$ is zero) for the epochs without data. The model $\hat{\tbf{s}}$ is constructed as before, but we then add a random component $\tbf{u}$ with the covariance matrix 
\begin{equation}
Q = (S^{-1} + N^{-1})^{-1}  .
\label{eq:Q}
\end{equation}
To construct $\tbf{u}$, we Cholesky decompose $Q = M^TM$, and the random component is simply $\tbf{u} = M\tbf{r}$, where $\tbf{r}$ is a vector of independent Gaussian random deviates with zero mean and unit standard deviation \citep[see][]{Zu2011}.

We fit the four observed lightcurves with {\tt JAVELIN}. The time baselines of these lightcurves are too short to well constrain the time scale $\tau_{\mbox{\tiny DRW}}$, so we fix $\tau_{\mbox{\tiny DRW}}$ to the estimated value from the empirical relation of \citet{MacLeod2010},
\begin{multline}
{\rm log}(\tau_{\mbox{\tiny DRW}}) = A + B\,{\rm log}(\lambda_{\rm RF}/4000\angstrom) \\
+ C (M_i + 23) + D\,{\rm log}(M_{\rm BH}/10^9 M_{\odot})
\label{eq:taudrw}
\end{multline}
where $\lambda_{\rm RF}$ is the rest-frame wavelength of the observation, $M_i$ is the \textit{i}-band absolute magnitude, $M_{\rm BH}$ is the black hole mass and $(A,B,C,D) = (2.4,0.17,0.03,0.21)$. We use the best-fit lightcurves and DRW parameters (when using the DRW model) to create simulated constrained lightcurves with 20 times the cadence of the observed lightcurves. We then resample the high-cadence simulated lightcurves to the cadence of the observed lightcurves through linear interpolations. Since the resampled lightcurves have much lower cadence than the original ones, the linear interpolation is adequate for the resampling. The exact value of $\tau_{\mbox{\tiny DRW}}$ is not critical to the results. For example, even if we fix $\tau_{\mbox{\tiny DRW}}$ in {\tt JAVELIN} to $1/10$ or 10 times the standard value, there is little effect on the lag estimates. Therefore, only a rough estimate of $\tau_{\mbox{\tiny DRW}}$ is needed. 

Given the mean noise of the observed lightcurve $\langle \sigma_i \rangle$, we add Gaussian noise of dispersion $X_i \langle \sigma_i \rangle$ to the simulated lightcurves, where the coefficient $X_i$ may depend on the epoch. For the analysis of the lightcurves, we say that the error is $Y_i \langle \sigma_i \rangle$, where this assigned uncertainty may differ from the actual noise (i.e., $X_i \neq Y_i$). We convolve the noiseless high-cadence continuum lightcurves with a transfer function $\Psi(\tau)$ to create emission-line lightcurves. The transfer function $\Psi(\tau)$ has a small random mean lag $t_0$ between 2 and 4 days. This is simply to produce random offsets between the continuum and the line measurement epochs. While these small lag values were originally motivated by the small continuum lags, the particular value of the lag is unimportant for our simulations and the discussions are equally applicable to both line RM and continuum RM. We resample and add noise to the line lightcurves following the same process as for the continuum.

We focus on constrained realizations of actual AGN lightcurves to avoid any concern that the model lightcurves are somehow not representative of real AGNs. We did carry out a full set of experiments with unconstrained random realizations of lightcurves, and they produce similar results to those we describe below. There is one easily understood difference. We know that the constrained realizations will yield well-defined lags since they are based on lightcurves chosen for analysis and publication because they yielded lags. What we are concerned with here is whether those lags are accurate in the sense that the estimated lag and its uncertainty are consistent with the true lag. 

Random lightcurve realizations with the same cadence and noise levels are not guaranteed to yield lags because sometimes the lightcurve has no significant features (i.e., curvature, maxima, minima) to allow a lag estimate. In such cases, any analysis will fail to give a significant lag measurement. As noted in the introduction, the probability that a given sampling strategy will yield a lightcurve that will produce a lag has been well-studied \citep[e.g.,][]{Horne2004,King2015,Shen2015,Yu2018,Li2019}, which is why we do not make it a focus of our study.

\section{Results} \label{sec:anl}

\begin{figure*}
\includegraphics[width=\linewidth]{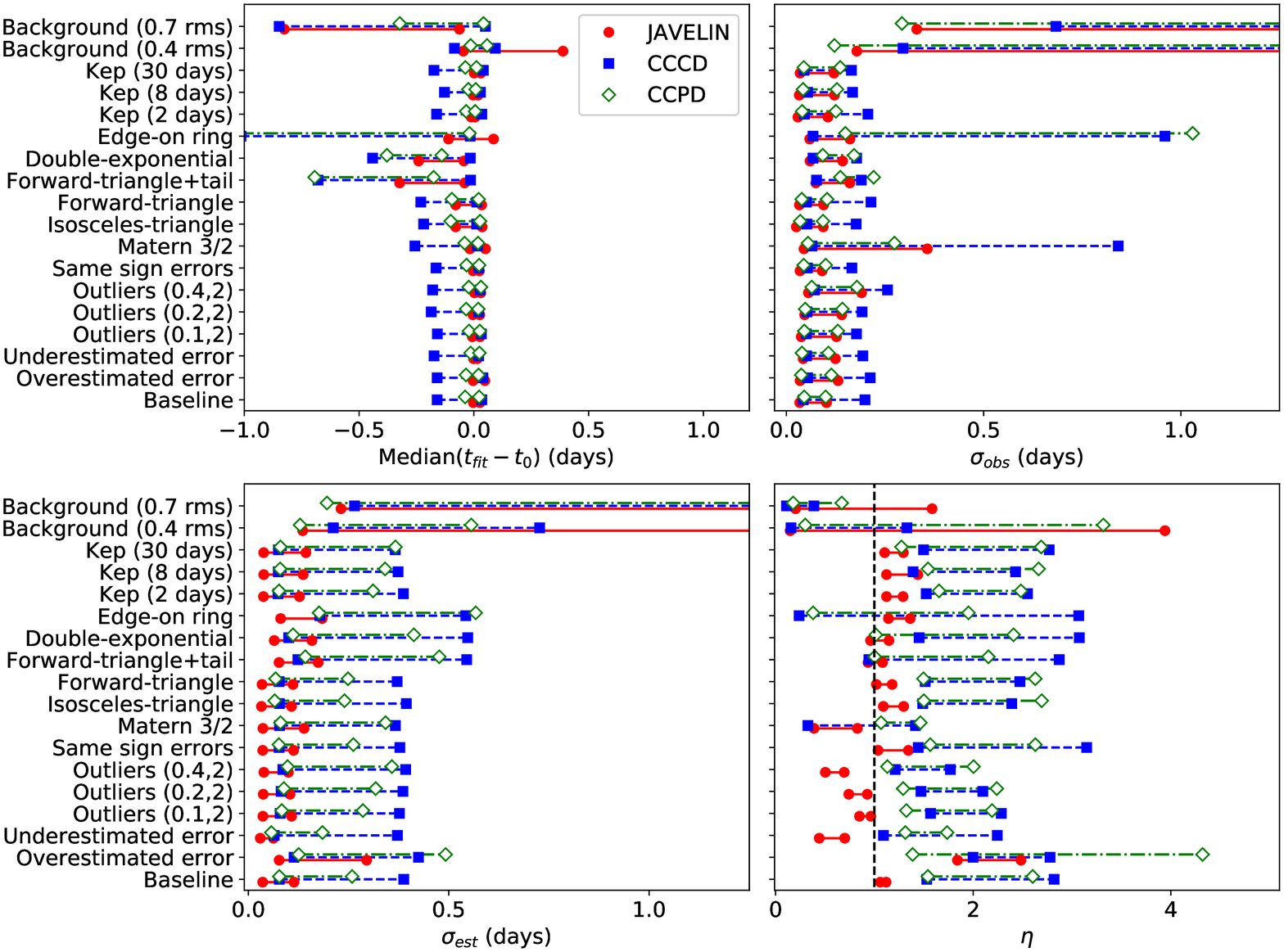}
\caption{Comparison of the median $(t_{\rm fit} - t_0)$ (upper left), $\sigma_{\rm obs}$ (upper right), $\sigma_{\rm est}$ (bottom left) and $\eta$ (bottom right) estimates for the different cases. For the background variation cases, we only include the two with changing random seeds for each realization. For each case, the red circles, blue squares and green empty diamonds are drawn at the maximum and the minimum parameter values for the four AGNs from {\tt JAVELIN}, CCCD and CCPD, respectively. We add small shifts along the y-axis for each case to avoid overlapping between the points and lines. In the bottom right panel, the black dashed line is drawn at $\eta=1$. The CCCD results for the median $(t_{\rm fit} - t_0)$ in Mrk 509 deviate significantly from the input, so we did not include those results in the upper left panel of the figure for visibility.}
\label{fig:tbpara}
\end{figure*}

We use {\tt JAVELIN} and PyCCF \citep{pyccf}, a python interface for the ICCF method, to measure the lags from the simulated lightcurves. For PyCCF, we create 8000 realizations with both flux randomization (FR) and random sub-sampling (RSS), and adopt the realizations with $r_{\rm peak}>0.5$ to compute the ICCF lag uncertainties, where $r_{\rm peak}$ is the peak value of each CCF. Nearly all realizations pass the $r_{\rm peak}$ cut. For each CCF, we use the region with $r>0.8r_{\rm peak}$ to calculate the centroid and the peak. We compare the input lag $t_0$ and the output lags $t_{\rm fit}$ and characterize the results by four parameters: (1) the median of $(t_{\rm fit} - t_0)$; (2) the width $\sigma_{\rm obs}$ of the $(t_{\rm fit} - t_0)$ distribution, defined as half the difference between the 16th and 84th percentile; (3) the mean $\sigma_{\rm est}$ of the algorithm error estimates calculated as half the difference between the 16th and 84th percentile of the JAVELIN posterior probability distribution, CCCD or CCPD for each realization; and (4) the ratio $\eta = \sigma_{\rm est} / \sigma_{\rm obs}$ between the estimated uncertainty $\sigma_{\rm est}$ and the observed scatter $\sigma_{\rm obs}$, where $\sigma_{\rm obs}$ is an estimate of the ``true'' uncertainty of the lag measurements and $\eta$ indicates whether the algorithms overestimate ($\eta>1$) or underestimate ($\eta<1$) the lag uncertainties. We show these parameters for all the cases we consider in Tables \ref{tab:results1} - \ref{tab:results_kepobs}. Figure \ref{fig:tbpara} illustrates the parameter ranges for the different cases we consider. 

\begin{table*}
\tiny
\renewcommand{\arraystretch}{1.06}
\begin{tabular}{llrrrrrrrrrrrr}
\hline
Configurations & Object & \multicolumn{3}{c}{Median $(t_{\rm fit} - t_0)$ (days)} & \multicolumn{3}{c}{$\sigma_{\rm obs}$ (days)} & \multicolumn{3}{c}{$\sigma_{\rm est}$ (days)} & \multicolumn{3}{c}{$\eta$} \\
 & & JAVELIN & CCCD & CCPD & JAVELIN & CCCD & CCPD & JAVELIN & CCCD & CCPD & JAVELIN & CCCD & CCPD\\
\hline
\multirow{4}{*}{Baseline}
& NGC 5548 & 0.026 & 0.035 & 0.023 & 0.046 & 0.065 & 0.052 & 0.051 & 0.184 & 0.080 & 1.11 & 2.82 & 1.54\\
& NGC 4151 & $-$0.002 & $-$0.160 & $-$0.038 & 0.102 & 0.168 & 0.099 & 0.114 & 0.257 & 0.259 & 1.12 & 1.53 & 2.60\\
& NGC 4593 & 0.000 & $-$0.110 & $-$0.013 & 0.034 & 0.043 & 0.045 & 0.036 & 0.077 & 0.077 & 1.06 & 1.81 & 1.71\\
& Mrk 509 & 0.017 & $-$1.207 & $-$0.055 & 0.095 & 0.199 & 0.089 & 0.104 & 0.388 & 0.159 & 1.09 & 1.95 & 1.80\\
\hline
\multirow{2}{*}{Overestimated}
& NGC 5548 & 0.011 & 0.039 & 0.021 & 0.046 & 0.079 & 0.091 & 0.102 & 0.220 & 0.126 & 2.24 & 2.78 & 1.39\\
\multirow{2}{*}{errors ($Y_i = 2$)}
& NGC 4151 & 0.041 & $-$0.160 & $-$0.034 & 0.119 & 0.157 & 0.114 & 0.295 & 0.410 & 0.492 & 2.48 & 2.61 & 4.32\\
& NGC 4593 & $-$0.003 & $-$0.132 & $-$0.025 & 0.035 & 0.054 & 0.038 & 0.077 & 0.114 & 0.132 & 2.21 & 2.12 & 3.51\\
& Mrk 509 & 0.048 & $-$1.303 & $-$0.101 & 0.131 & 0.212 & 0.094 & 0.241 & 0.425 & 0.297 & 1.84 & 2.00 & 3.18\\
\hline
\multirow{2}{*}{Underestimated}
& NGC 5548 & 0.014 & 0.021 & 0.024 & 0.064 & 0.077 & 0.052 & 0.037 & 0.172 & 0.068 & 0.59 & 2.24 & 1.31\\
\multirow{2}{*}{(errors $Y_i = 0.5$)}
& NGC 4151 & 0.015 & $-$0.174 & $-$0.007 & 0.124 & 0.179 & 0.107 & 0.063 & 0.196 & 0.185 & 0.50 & 1.09 & 1.74\\
& NGC 4593 & $-$0.001 & $-$0.110 & $-$0.014 & 0.043 & 0.051 & 0.039 & 0.030 & 0.065 & 0.057 & 0.70 & 1.26 & 1.46\\
& Mrk 509 & 0.013 & $-$1.198 & $-$0.019 & 0.119 & 0.194 & 0.087 & 0.053 & 0.372 & 0.126 & 0.44 & 1.92 & 1.45\\
\hline
\multirow{2}{*}{Outliers}
& NGC 5548 & 0.027 & 0.033 & 0.026 & 0.054 & 0.081 & 0.064 & 0.051 & 0.184 & 0.084 & 0.96 & 2.28 & 1.32\\
\multirow{2}{*}{($f_{out} = 0.1$)}
& NGC 4151 & 0.007 & $-$0.159 & $-$0.017 & 0.127 & 0.166 & 0.130 & 0.108 & 0.266 & 0.286 & 0.85 & 1.60 & 2.19\\
& NGC 4593 & $-$0.007 & $-$0.124 & $-$0.021 & 0.038 & 0.051 & 0.045 & 0.036 & 0.079 & 0.083 & 0.96 & 1.57 & 1.83\\
& Mrk 509 & 0.019 & $-$1.267 & $-$0.056 & 0.113 & 0.178 & 0.093 & 0.102 & 0.377 & 0.180 & 0.90 & 2.12 & 1.94\\
\hline
\multirow{2}{*}{Outliers}
& NGC 5548 & 0.020 & 0.021 & 0.019 & 0.057 & 0.089 & 0.071 & 0.053 & 0.186 & 0.091 & 0.93 & 2.10 & 1.29\\
\multirow{2}{*}{($f_{out} = 0.2$)}
& NGC 4151 & 0.006 & $-$0.186 & $-$0.034 & 0.140 & 0.188 & 0.142 & 0.104 & 0.276 & 0.318 & 0.74 & 1.47 & 2.24\\
& NGC 4593 & $-$0.004 & $-$0.123 & $-$0.021 & 0.046 & 0.052 & 0.048 & 0.037 & 0.082 & 0.088 & 0.80 & 1.56 & 1.86\\
& Mrk 509 & 0.026 & $-$1.222 & $-$0.054 & 0.117 & 0.192 & 0.119 & 0.099 & 0.386 & 0.184 & 0.84 & 2.01 & 1.54\\
\hline
\multirow{2}{*}{Outliers}
& NGC 5548 & 0.017 & 0.028 & 0.031 & 0.094 & 0.109 & 0.087 & 0.053 & 0.192 & 0.098 & 0.57 & 1.77 & 1.13\\
\multirow{2}{*}{($f_{out} = 0.4$)}
& NGC 4151 & 0.013 & $-$0.179 & $-$0.017 & 0.187 & 0.216 & 0.179 & 0.094 & 0.298 & 0.358 & 0.50 & 1.38 & 2.00\\
& NGC 4593 & 0.003 & $-$0.132 & $-$0.022 & 0.056 & 0.071 & 0.064 & 0.039 & 0.086 & 0.100 & 0.69 & 1.21 & 1.55\\
& Mrk 509 & 0.030 & $-$1.306 & $-$0.085 & 0.191 & 0.256 & 0.146 & 0.100 & 0.393 & 0.213 & 0.52 & 1.53 & 1.46\\
\hline
\multirow{2}{*}{Correlated errors}
& NGC 5548 & 0.024 & 0.021 & 0.023 & 0.045 & 0.058 & 0.051 & 0.051 & 0.182 & 0.080 & 1.14 & 3.15 & 1.57\\
\multirow{2}{*}{(Same sign)}
& NGC 4151 & 0.004 & $-$0.165 & $-$0.032 & 0.091 & 0.127 & 0.100 & 0.113 & 0.252 & 0.262 & 1.25 & 1.98 & 2.63\\
& NGC 4593 & $-$0.003 & $-$0.115 & $-$0.014 & 0.035 & 0.053 & 0.044 & 0.036 & 0.077 & 0.076 & 1.04 & 1.44 & 1.74\\
& Mrk 509 & 0.016 & $-$1.189 & $-$0.082 & 0.078 & 0.166 & 0.093 & 0.105 & 0.378 & 0.174 & 1.34 & 2.28 & 1.87\\
\hline
\multirow{2}{*}{Correlated errors}
& NGC 5548 & 0.012 & 0.016 & 0.018 & 0.065 & 0.130 & 0.072 & 0.054 & 0.184 & 0.088 & 0.83 & 1.42 & 1.22\\
\multirow{2}{*}{(Matern 3/2)}
& NGC 4151 & $-$0.017 & $-$0.257 & $-$0.040 & 0.357 & 0.841 & 0.274 & 0.139 & 0.275 & 0.342 & 0.39 & 0.33 & 1.25\\
& NGC 4593 & $-$0.003 & $-$0.120 & $-$0.018 & 0.045 & 0.065 & 0.055 & 0.036 & 0.078 & 0.080 & 0.82 & 1.20 & 1.47\\
& Mrk 509 & 0.051 & $-$1.261 & $-$0.093 & 0.175 & 0.388 & 0.166 & 0.106 & 0.367 & 0.177 & 0.60 & 0.95 & 1.07\\
\hline
\end{tabular}
\caption{Simulation results for different sources and model configurations. Column (1) and (2) give the configuration descriptions and the object names, respectively. Column (3) - (5) give the median of $(t_{\rm fit} - t_0)$ from {\tt JAVELIN}, CCCD and CCPD, respectively, where $t_{0}$ is the input lag and $t_{\rm fit}$ is the best-fit lag from the algorithms. Column (6) - (8) give the scatter $\sigma_{\rm obs}$ in $(t_{\rm fit} - t_0)$, defined as half the difference between the 16th and 84th percentile. Column (9) - (11) give the mean error estimates $\sigma_{\rm est}$ from the algorithms. Column (12) - (14) give the ratio $\eta = \sigma_{\rm est} / \sigma_{\rm obs}$, where $\eta>1$ ($\eta<1$) means that the lag uncertainties are over (under) estimated.}
\label{tab:results1}
\end{table*}

\begin{table*}
\tiny
\renewcommand{\arraystretch}{1.06}
\begin{tabular}{llrrrrrrrrrrrr}
\hline
Configurations & Object & \multicolumn{3}{c}{Median $(t_{\rm fit} - t_0)$ (days)} & \multicolumn{3}{c}{$\sigma_{\rm obs}$ (days)} & \multicolumn{3}{c}{$\sigma_{\rm est}$ (days)} & \multicolumn{3}{c}{$\eta$} \\
 & & JAVELIN & CCCD & CCPD & JAVELIN & CCCD & CCPD & JAVELIN & CCCD & CCPD & JAVELIN & CCCD & CCPD\\
\hline
\multirow{2}{*}{Isosceles-}
& NGC 5548 & 0.003 & 0.012 & 0.011 & 0.045 & 0.077 & 0.055 & 0.049 & 0.184 & 0.082 & 1.09 & 2.39 & 1.50\\
\multirow{2}{*}{triangular $\Psi(\tau)$}
& NGC 4151 & $-$0.080 & $-$0.219 & $-$0.101 & 0.087 & 0.159 & 0.089 & 0.108 & 0.258 & 0.240 & 1.24 & 1.62 & 2.69\\
& NGC 4593 & 0.036 & $-$0.074 & 0.027 & 0.025 & 0.052 & 0.035 & 0.033 & 0.077 & 0.066 & 1.30 & 1.49 & 1.88\\
& Mrk 509 & 0.033 & $-$1.176 & $-$0.030 & 0.093 & 0.177 & 0.093 & 0.103 & 0.394 & 0.158 & 1.10 & 2.23 & 1.71\\
\hline
\multirow{2}{*}{Forward-}
& NGC 5548 & 0.002 & 0.013 & 0.007 & 0.049 & 0.074 & 0.054 & 0.050 & 0.184 & 0.081 & 1.02 & 2.47 & 1.50\\
\multirow{2}{*}{triangular $\Psi(\tau)$}
& NGC 4151 & $-$0.079 & $-$0.231 & $-$0.096 & 0.094 & 0.167 & 0.095 & 0.111 & 0.255 & 0.249 & 1.18 & 1.52 & 2.63\\
& NGC 4593 & 0.032 & $-$0.078 & 0.021 & 0.033 & 0.051 & 0.039 & 0.034 & 0.077 & 0.068 & 1.03 & 1.51 & 1.75\\
& Mrk 509 & 0.034 & $-$1.228 & $-$0.031 & 0.090 & 0.214 & 0.103 & 0.102 & 0.371 & 0.157 & 1.13 & 1.73 & 1.52\\
\hline
\multirow{4}{*}{Long-tail $\Psi(\tau)$}
& NGC 5548 & $-$0.110 & $-$0.014 & $-$0.175 & 0.075 & 0.076 & 0.142 & 0.077 & 0.177 & 0.142 & 1.03 & 2.32 & 1.00\\
& NGC 4151 & $-$0.162 & $-$0.307 & $-$0.443 & 0.161 & 0.190 & 0.221 & 0.174 & 0.260 & 0.476 & 1.08 & 1.37 & 2.15\\
& NGC 4593 & $-$0.323 & $-$0.679 & $-$0.694 & 0.105 & 0.131 & 0.137 & 0.098 & 0.124 & 0.210 & 0.94 & 0.94 & 1.53\\
& Mrk 509 & $-$0.040 & $-$0.248 & $-$0.109 & 0.133 & 0.190 & 0.149 & 0.137 & 0.545 & 0.264 & 1.03 & 2.87 & 1.77\\
\hline
\multirow{2}{*}{Double-}
& NGC 5548 & $-$0.136 & $-$0.016 & $-$0.140 & 0.060 & 0.067 & 0.110 & 0.069 & 0.178 & 0.111 & 1.15 & 2.64 & 1.01\\
\multirow{2}{*}{exponential $\Psi(\tau)$}
& NGC 4151 & $-$0.200 & $-$0.306 & $-$0.350 & 0.143 & 0.157 & 0.171 & 0.159 & 0.256 & 0.413 & 1.11 & 1.63 & 2.41\\
& NGC 4593 & $-$0.240 & $-$0.441 & $-$0.379 & 0.067 & 0.069 & 0.092 & 0.065 & 0.101 & 0.163 & 0.96 & 1.45 & 1.77\\
& Mrk 509 & $-$0.043 & $-$0.269 & $-$0.093 & 0.118 & 0.178 & 0.105 & 0.131 & 0.547 & 0.233 & 1.11 & 3.07 & 2.21\\
\hline
\multirow{4}{*}{Edge-on ring's $\Psi(\tau)$}
& NGC 5548 & $-$0.048 & $-$0.016 & $-$0.019 & 0.059 & 0.068 & 0.163 & 0.081 & 0.178 & 0.177 & 1.36 & 2.64 & 1.09\\
& NGC 4151 & $-$0.111 & $-$0.348 & $-$0.324 & 0.161 & 0.187 & 0.291 & 0.184 & 0.277 & 0.567 & 1.14 & 1.48 & 1.95\\
& NGC 4593 & 0.086 & $-$1.014 & $-$1.549 & 0.131 & 0.959 & 1.030 & 0.152 & 0.229 & 0.391 & 1.16 & 0.24 & 0.38\\
& Mrk 509 & $-$0.025 & $-$0.294 & $-$0.077 & 0.107 & 0.177 & 0.149 & 0.142 & 0.543 & 0.289 & 1.32 & 3.07 & 1.93\\
\hline
\multirow{2}{*}{``Kepler'' process}
& NGC 5548 & 0.002 & 0.035 & 0.005 & 0.043 & 0.069 & 0.051 & 0.055 & 0.175 & 0.084 & 1.28 & 2.55 & 1.65\\
\multirow{2}{*}{($\tau_c = 2 \,{\rm days}$)}
& NGC 4151 & $-$0.011 & $-$0.163 & $-$0.034 & 0.105 & 0.165 & 0.125 & 0.128 & 0.251 & 0.311 & 1.21 & 1.52 & 2.48\\
& NGC 4593 & $-$0.002 & $-$0.113 & $-$0.014 & 0.029 & 0.046 & 0.040 & 0.038 & 0.074 & 0.077 & 1.29 & 1.61 & 1.92\\
& Mrk 509 & 0.005 & $-$1.210 & $-$0.094 & 0.102 & 0.207 & 0.096 & 0.115 & 0.387 & 0.179 & 1.12 & 1.87 & 1.86\\
\hline
\multirow{2}{*}{``Kepler'' process}
& NGC 5548 & $-$0.003 & 0.029 & 0.008 & 0.040 & 0.071 & 0.056 & 0.057 & 0.172 & 0.086 & 1.44 & 2.43 & 1.54\\
\multirow{2}{*}{($\tau_c = 8 \,{\rm days}$)}
& NGC 4151 & 0.007 & $-$0.128 & $-$0.024 & 0.122 & 0.168 & 0.128 & 0.137 & 0.246 & 0.341 & 1.12 & 1.47 & 2.66\\
& NGC 4593 & 0.001 & $-$0.114 & $-$0.006 & 0.033 & 0.054 & 0.042 & 0.038 & 0.075 & 0.079 & 1.18 & 1.39 & 1.88\\
& Mrk 509 & 0.018 & $-$1.173 & $-$0.067 & 0.106 & 0.165 & 0.102 & 0.124 & 0.374 & 0.187 & 1.17 & 2.26 & 1.84\\
\hline
\multirow{2}{*}{``Kepler'' process}
& NGC 5548 & 0.002 & 0.041 & 0.011 & 0.048 & 0.069 & 0.064 & 0.058 & 0.170 & 0.081 & 1.19 & 2.47 & 1.27\\
\multirow{2}{*}{($\tau_c = 30 \,{\rm days}$)}
& NGC 4151 & 0.016 & $-$0.174 & $-$0.037 & 0.121 & 0.165 & 0.137 & 0.144 & 0.247 & 0.367 & 1.19 & 1.50 & 2.69\\
& NGC 4593 & 0.002 & $-$0.111 & $-$0.009 & 0.035 & 0.045 & 0.044 & 0.038 & 0.075 & 0.080 & 1.10 & 1.66 & 1.81\\
& Mrk 509 & 0.030 & $-$1.204 & $-$0.058 & 0.099 & 0.132 & 0.081 & 0.127 & 0.366 & 0.188 & 1.29 & 2.77 & 2.33\\
\hline
\end{tabular}
\caption{Same as Table \ref{tab:results1} but for different model configurations.}
\label{tab:results2}
\end{table*}

\begin{table*}
\tiny
\renewcommand{\arraystretch}{1.06}
\begin{tabular}{lccrrrrrrrrrrrr}
\hline
Object & Seed & $\sigma_{\rm bkg}$ & \multicolumn{3}{c}{Median $(t_{\rm fit} - t_0)$ (days)} & \multicolumn{3}{c}{$\sigma_{\rm obs}$ (days)} & \multicolumn{3}{c}{$\sigma_{\rm est}$ (days)} & \multicolumn{3}{c}{$\eta$} \\
 & & & JAVELIN & CCCD & CCPD & JAVELIN & CCCD & CCPD & JAVELIN & CCCD & CCPD & JAVELIN & CCCD & CCPD\\
\hline
\multirow{6}{*}{NGC 5548}
& 20 & 3.430 & 0.206 & 0.530 & 0.250 & 0.123 & 0.075 & 0.078 & 0.127 & 0.209 & 0.136 & 1.04 & 2.78 & 1.75 \\
& 20 & 6.002 & 0.430 & 0.867 & 0.401 & 0.249 & 0.096 & 0.139 & 0.239 & 0.264 & 0.211 & 0.96 & 2.74 & 1.51 \\
& 30 & 3.430 & 0.030 & 0.189 & 0.081 & 0.130 & 0.071 & 0.057 & 0.140 & 0.221 & 0.127 & 1.07 & 3.12 & 2.24 \\
& 30 & 6.002 & $-$0.067 & 0.335 & 0.137 & 0.215 & 0.092 & 0.082 & 0.232 & 0.274 & 0.195 & 1.08 & 2.98 & 2.38 \\
& random & 3.430 & 0.019 & 0.066 & 0.057 & 0.179 & 0.375 & 0.166 & 0.136 & 0.212 & 0.129 & 0.76 & 0.56 & 0.78 \\
& random & 6.002 & $-$0.063 & 0.050 & 0.042 & 0.330 & 0.683 & 0.293 & 0.231 & 0.265 & 0.196 & 0.70 & 0.39 & 0.67 \\
\hline
\multirow{6}{*}{NGC 4151}
& 40 & 0.150 & 0.929 & 1.121 & 0.520 & 0.293 & 0.212 & 0.235 & 0.167 & 0.263 & 0.420 & 0.57 & 1.24 & 1.79 \\
& 40 & 0.262 & 1.877 & 1.909 & 1.283 & 0.330 & 0.267 & 0.363 & 0.170 & 0.284 & 0.586 & 0.52 & 1.07 & 1.61 \\
& 50 & 0.150 & 0.074 & 0.556 & 0.059 & 0.152 & 0.232 & 0.140 & 0.118 & 0.315 & 0.343 & 0.78 & 1.36 & 2.46 \\
& 50 & 0.262 & 0.167 & 1.332 & 0.082 & 0.430 & 0.660 & 0.275 & 0.447 & 0.674 & 0.571 & 1.04 & 1.02 & 2.08 \\
& random & 0.150 & 0.139 & 0.093 & 0.034 & 0.775 & 1.575 & 0.628 & 0.151 & 0.306 & 0.429 & 0.19 & 0.19 & 0.68 \\
& random & 0.262 & $-$0.077 & $-$0.393 & $-$0.134 & 2.134 & 3.305 & 1.857 & 0.984 & 0.502 & 0.704 & 0.46 & 0.15 & 0.38 \\
\hline
\multirow{6}{*}{NGC 4593}
& 60 & 0.064 & $-$5.631 & $-$0.298 & $-$0.060 & 3.351 & 0.099 & 0.054 & 3.728 & 0.108 & 0.109 & 1.11 & 1.09 & 2.03 \\
& 60 & 0.111 & $-$5.940 & $-$6.012 & $-$6.058 & 0.557 & 1.447 & 0.690 & 3.321 & 0.914 & 1.063 & 5.97 & 0.63 & 1.54 \\
& 70 & 0.064 & 0.100 & 0.157 & 0.106 & 0.109 & 0.058 & 0.065 & 6.424 & 0.861 & 0.891 & 58.89 & 14.87 & 13.79 \\
& 70 & 0.111 & 0.101 & $-$16.176 & $-$16.244 & 3.458 & 8.607 & 8.586 & 7.823 & 3.273 & 3.343 & 2.26 & 0.38 & 0.39 \\
& random & 0.064 & $-$0.046 & $-$0.085 & $-$0.014 & 1.307 & 0.295 & 0.122 & 5.151 & 0.393 & 0.404 & 3.94 & 1.33 & 3.32 \\
& random & 0.111 & $-$0.826 & $-$0.848 & $-$0.323 & 3.177 & 7.787 & 7.906 & 5.034 & 1.357 & 1.409 & 1.58 & 0.17 & 0.18 \\
\hline
\multirow{6}{*}{Mrk 509}
& 80 & 0.417 & 2.257 & 6.766 & 2.159 & 0.444 & 0.340 & 0.264 & 0.305 & 0.842 & 0.634 & 0.69 & 2.48 & 2.40 \\
& 80 & 0.729 & 3.597 & 21.705 & 17.775 & 0.973 & 0.516 & 11.641 & 0.689 & 1.023 & 13.166 & 0.71 & 1.98 & 1.13 \\
& 90 & 0.417 & $-$3.480 & $-$4.502 & $-$1.037 & 0.765 & 0.364 & 0.350 & 0.605 & 0.951 & 0.876 & 0.79 & 2.61 & 2.50 \\
& 90 & 0.729 & $-$7.618 & $-$8.109 & $-$6.199 & 1.941 & 0.250 & 0.543 & 1.390 & 0.888 & 2.170 & 0.72 & 3.55 & 3.99 \\
& random & 0.417 & 0.388 & $-$0.563 & $-$0.027 & 2.545 & 4.658 & 1.850 & 0.377 & 0.727 & 0.556 & 0.15 & 0.16 & 0.30 \\
& random & 0.729 & $-$0.262 & $-$0.395 & 0.029 & 3.944 & 9.468 & 3.999 & 0.800 & 1.055 & 1.255 & 0.20 & 0.11 & 0.31 \\
\hline
\end{tabular}
\caption{Simulation results from varying the backgrounds of the line lightcurves. Column (2) gives the random seed used to generate the background variation. The notation ``random'' means we used a different random seed for each realization. Column (3) gives the standard deviation $\sigma_{\rm bkg}$ of the background variation. Other columns have the same meaning as Table \ref{tab:results1}.}
\label{tab:results_bkg}
\end{table*}

\begin{table*}
\tiny
\renewcommand{\arraystretch}{1.06}
\begin{tabular}{lccrrrrrrrrrrrr}
\hline
Cadence & Time Interval & \multicolumn{3}{c}{Median $(t_{\rm fit} - t_0)$ (days)} & \multicolumn{3}{c}{$\sigma_{\rm obs}$ (days)} & \multicolumn{3}{c}{$\sigma_{\rm est}$ (days)} & \multicolumn{3}{c}{$\eta$} \\
 & (HJD $-$ 2400000) & JAVELIN & CCCD & CCPD & JAVELIN & CCCD & CCPD & JAVELIN & CCCD & CCPD & JAVELIN & CCCD & CCPD\\
\hline
\multirow{2}{*}{NGC 4151}
& (55641.515 , 55710.915) & $-$0.000 & $-$0.038 & $-$0.064 & 0.134 & 0.296 & 0.174 & 0.182 & 0.514 & 0.440 & 1.36 & 1.74 & 2.53 \\
\multirow{2}{*}{Cadence}
& (56321.668 , 56391.068) & 0.019 & 0.051 & 0.021 & 0.105 & 0.125 & 0.137 & 0.123 & 0.204 & 0.314 & 1.17 & 1.64 & 2.30 \\
& (55757.235 , 55826.635) & 0.030 & $-$0.719 & $-$0.037 & 0.187 & 0.541 & 0.220 & 0.208 & 0.889 & 0.535 & 1.11 & 1.64 & 2.43 \\
& (56134.419 , 56203.819) & $-$0.000 & 0.084 & 0.005 & 0.057 & 0.098 & 0.072 & 0.069 & 0.134 & 0.163 & 1.21 & 1.37 & 2.26 \\
\hline
\multirow{2}{*}{NGC 4593}
& (55400.383 , 55423.083) & 0.009 & $-$0.445 & $-$0.018 & 0.072 & 0.102 & 0.081 & 0.083 & 0.205 & 0.193 & 1.14 & 2.02 & 2.39 \\
\multirow{2}{*}{Cadence}
& (55871.416 , 55894.116) & 0.015 & $-$0.277 & $-$0.017 & 0.092 & 0.393 & 0.111 & 0.112 & 0.448 & 0.247 & 1.22 & 1.14 & 2.23 \\
& (56210.391 , 56233.091) & 0.065 & $-$0.616 & $-$0.108 & 0.093 & 0.082 & 0.113 & 0.106 & 0.175 & 0.242 & 1.14 & 2.14 & 2.15 \\
& (56277.349 , 56300.049) & 0.019 & $-$0.230 & $-$0.055 & 0.068 & 0.081 & 0.084 & 0.076 & 0.128 & 0.175 & 1.12 & 1.58 & 2.08 \\
\hline
\end{tabular}
\caption{Results from simulated lightcurves based on the Kepler lightcurves of Zw 229$-$15. The Kepler lightcurves are sampled at the cadence of the Swift lightcurves of either NGC 4151 or NGC 4593. Column (2) gives the start and end time of the time intervals where we sample the Kepler lightcurve, respectively. Other columns have the same meaning as Table \ref{tab:results1}.}
\label{tab:results_kepobs}
\end{table*}

\subsection{Baseline Configuration} \label{subsec:anl_ini}
We first create simulated lightcurves that satisfy all the assumptions made by {\tt JAVELIN}. We adopted DRW models with the parameters $(\sigma_{\mbox{\tiny DRW}},\tau_{\mbox{\tiny DRW}}) = (17.38,125)\, ,\, (0.72,136)\, ,\, (0.49,86)\, ,\, (0.77,146)$ for NGC 5548, NGC 4151, NGC 4593 and Mrk 509, respectively, where $\tau_{\mbox{\tiny DRW}}$ is in units of days and $\sigma_{\mbox{\tiny DRW}}$ is in the same flux units as the observed lightcurves. We created 200 realizations of the simulated continuum for each object. For each realization, we construct the line lightcurves by convolving the simulated continuum with a normalized top-hat transfer function with a width of 0.6 days and a random lag between 2 days and 4 days. The particular value of the lag is unimportant here. We use some spread so that the alignment of the continuum and the line epochs varies. We assume Gaussian uncorrelated and correctly estimated noise like that in the observations (i.e., $X_i \equiv Y_i \equiv 1$). We then estimated the lags for all 800 lightcurves with both {\tt JAVELIN} and ICCF. Several {\tt JAVELIN} lag distributions for NGC 4593 show a weak secondary peak at $\sim -10$ days due to aliasing. Since this effect is well-understood, we only consider the lag distribution between $-$2 days to 8 days for the uncertainty estimates in our analysis.

\begin{figure*}
\includegraphics[width=\linewidth]{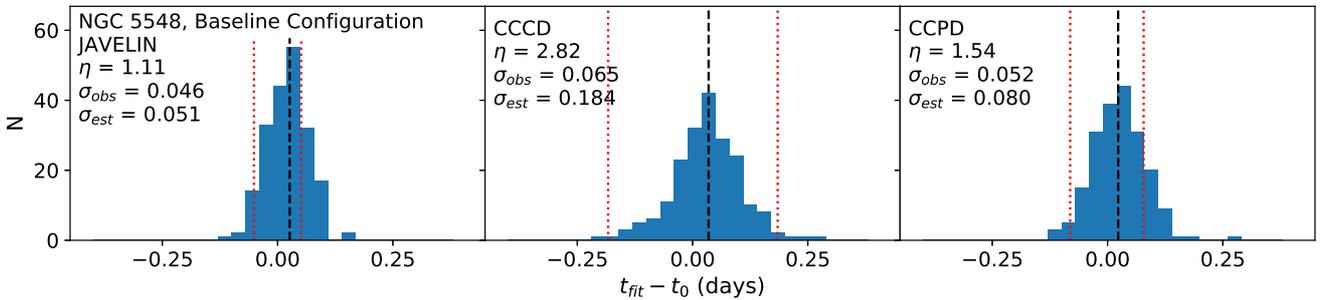}
\caption{Distribution of the difference between the best-fit lags $t_{\rm fit}$ and the input lags $t_0$ for NGC 5548. The black dashed lines are drawn at the median of the distribution $(t_{\rm fit}-t_0)$, while the red dotted lines are drawn at the mean of the 1$\sigma$ upper and lower limits estimated by the algorithms. The upper left corner of each panel reports the scatter $\sigma_{\rm obs}$ in the $(t_{\rm fit}-t_0)$ distributions, the mean error estimates $\sigma_{\rm est}$ from the algorithms and the ratio $\eta = \sigma_{\rm est} / \sigma_{\rm obs}$. The left, middle and right panel show the results from {\tt JAVELIN}, CCCD and CCPD, respectively.}
\label{fig:anl_ini}
\end{figure*}

Figure \ref{fig:anl_ini} shows the distribution of the difference $(t_{\rm fit} - t_0)$ between the best-fit lags $t_{\rm fit}$ and the input lags $t_0$ for NGC 5548. We only show the results for NGC 5548 as an example in the main body of the paper and include the results for the other three objects in the online journal. The median of the $(t_{\rm fit}-t_0)$ distributions from all algorithms shows a slight offset from zero by around 0.02 days. This is likely a small artifact from the sampling or convolution process used to produce the simulated lightcurves. However, the 0.02 days offset is small compared the lag uncertainties and will not affect our conclusions. {\tt JAVELIN} gives the smallest scatter $\sigma_{\rm obs}$ and the smallest error estimates $\sigma_{\rm est}$ among the three distributions, while CCCD gives the largest $\sigma_{\rm obs}$ and $\sigma_{\rm est}$. All algorithms overestimate the lag uncertainties with $\eta>1$. The {\tt JAVELIN} lag uncertainties are closest to the ``true'' uncertainty with $\eta \approx 1.1$, while the CCCD and CCPD methods overestimate the lag uncertainties with $\eta \approx 3$ and $\eta \approx 1.5$, respectively. We briefly explored restricting the ICCF method to only FR or only RSS rather than both. In most cases this reduced the ratio $\eta$, but not in any systematic pattern, with FR sometimes having the greater effect and other times RSS. The ICCF method can underestimate the lag uncertainties (i.e., $\eta<1$) with only FR or only RSS for some cases, while still overestimate the uncertainties (i.e., $\eta>1$) for the others.

The other three objects generally show similar results to NGC 5548. The only poorly estimated lags, in the sense that the medians of $(t_{\rm fit} - t_0)$ are more than $2\sigma_{\rm est}$, are the CCCD estimates for Mrk 509 with a median $(t_{\rm fit} - t_0) = - 1.2$ days and $\sigma_{\rm est} = 0.38$ days. {\tt JAVELIN} consistently comes closest to correctly estimating the lag uncertainties with $\eta \approx 1.1$, while the CCCD and CCPD methods overestimate the uncertainties with $\eta$ from 1.5 to 2.8. CCCD overestimates the lag uncertainty more severely than CCPD for some objects, while CCPD performs worse for the others. These differences between the uncertainty estimates from JAVELIN and ICCF are similar to those found in real RM campaigns \citep[e.g.,][]{Fausnaugh2017,McHardy2018,Edelson2019}.

We take these results as a ``baseline'' for comparison with other cases. For the observed scatter $\sigma_{\rm obs}$, the estimated uncertainty $\sigma_{\rm est}$ and the ratio $\eta$, we say the parameter differs ``significantly'' from the baseline if the parameter changes by more than 25\%. For the median $(t_{\rm fit} - t_0)$, we do not say a change is significant as long as its absolute value is less than 0.1 days. We focus on the bulk behaviour in each case and do not discuss the behaviour of the individual objects in detail unless the results are driven by particular lightcurve features.

\subsection{Effect of Input Errors} \label{subsec:anl_errin}

In real RM campaigns, the uncertainties in the lightcurves may not be correctly estimated due to, for example, seeing-induced aperture effects on the spectra \citep[e.g.,][]{Peterson1995}. This will have consequences for the lag uncertainties. We consider several potential problems with the single-epoch error estimates. 

\subsubsection{Incorrect Error Estimates} \label{sssec:anl_errvar}
We first artificially overestimate or underestimate the single-epoch uncertainties. The lightcurves are unchanged from the baseline configuration, but we either double ($Y_i = 2$) or halve ($Y_i = 0.5$) the uncertainties assigned to both the continuum and the line lightcurves while keeping the actual noise unchanged ($X_i = 1$). That is, we feed the algorithms with single-epoch errors that are two times larger or smaller than the noise that was actually added to the lightcurves. 

\begin{figure*}
\includegraphics[width=\linewidth]{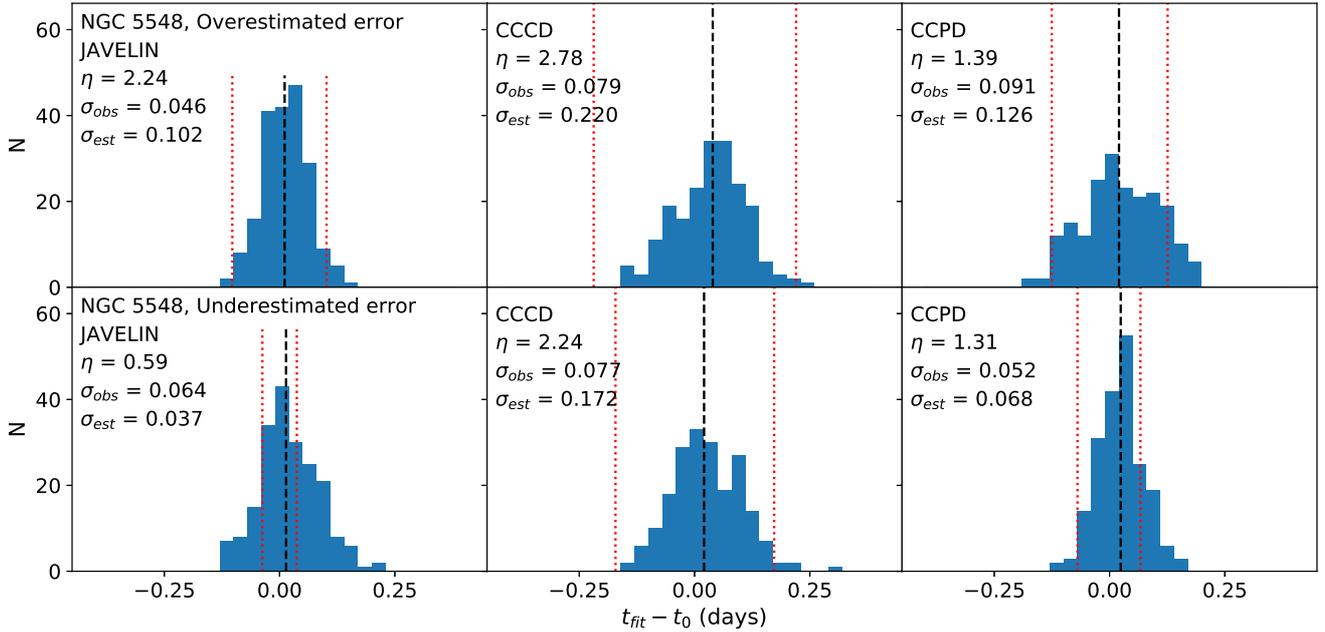}
\caption{Same as Figure \ref{fig:anl_ini} but for different configurations. The top row shows the results from overestimating the single-epoch errors, while the bottom row shows the results from underestimating the single-epoch errors.}
\label{fig:anl_errvar}
\end{figure*}

We show the results in Figure \ref{fig:anl_errvar}. Since the changes in this case have no effect on the ``shape'' of the lightcurves, it is not surprising to find only small differences in the median $(t_{\rm fit}-t_0)$ and the observed scatter $\sigma_{\rm obs}$ from the baseline case. When overestimating the uncertainties, the observed scatter $\sigma_{\rm obs}$ only varies slightly, except for the {\tt JAVELIN} results for Mrk 509 and the CCPD results for NGC 5548. When underestimating the uncertainties, $\sigma_{\rm obs}$ consistently increases for {\tt JAVELIN} while it changes little for CCCD and CCPD. Both algorithms give larger lag uncertainties when overestimating the single-epoch uncertainties and smaller lag uncertainties while underestimating the single-epoch uncertainties. The ratio $\eta$ roughly doubles/halves for {\tt JAVELIN} when we double/halve the uncertainties, as expected from its strong assumption of Gaussian $\chi^2$ uncertainties. On the other hand, the change in $\eta$ for CCCD and CCPD is generally smaller, and $\eta$ even slightly drops rather than increases when overestimating the uncertainties for NGC 5548. The ICCF method does not directly use the single-epoch uncertainties, which makes it less sensitive to incorrect estimates of the single-epoch uncertainties, albeit at the price of significantly overestimating the lag uncertainties if the error estimates are correct. 

\subsubsection{Outliers} \label{sssec:anl_outlier}
Rather than having incorrect error estimates for all epochs, a lightcurve can contain ``outliers'' that have intrinsically larger scatter than estimated. To simulate this, we select $f_{\rm out}N_{\rm p}$ points to be ``outliers'' in both the continuum and the line lightcurves, where $N_{\rm p}$ is the total number of epochs and $f_{\rm out}$ is the outlier fraction. We increase the intrinsic scatter of each outlier to $X_i = 2$ while keeping the intrinsic scatter of all other epochs at $X_i = 1$. While we tried outliers with larger scatters (e.g., $X_i = 8$), those outliers generally stand out from the lightcurves and can easily be identified and removed, so we do not consider those cases. We keep the assigned uncertainties unchanged for all epochs ($Y_i = 1$) so that the algorithms assume there are no outliers, and we consider outlier fractions of $f_{\rm out} = 0.1,\, 0.2,\, 0.4$. 

\begin{figure*}
\includegraphics[width=\linewidth]{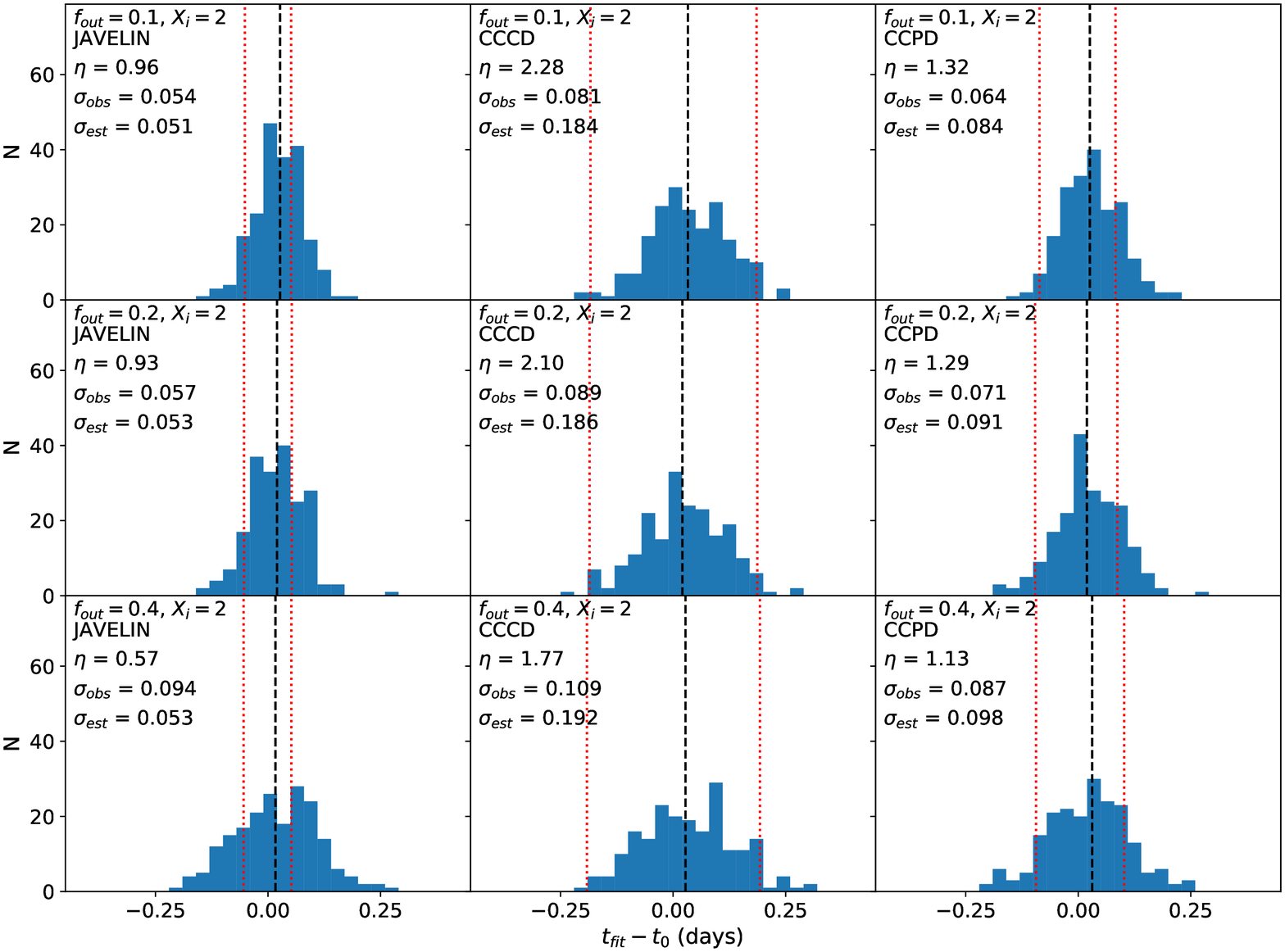}
\caption{Same as Figure \ref{fig:anl_ini}, but for the lightcurves with outliers for NGC 5548. The left, middle and right column show the results from {\tt JAVELIN}, CCCD and CCPD, respectively. The top, middle and bottom row show the results from $f_{\rm out} = 0.1,\, 0.2,\, 0.4$, respectively, where $f_{\rm out}$ is the fraction of the outliers.}
\label{fig:anl_outlierjav}
\end{figure*}

Figure \ref{fig:anl_outlierjav} shows the results for NGC 5548. The medians of $(t_{\rm fit}-t_0)$ show only small changes for both algorithms. The scatter $\sigma_{\rm obs}$ consistently increases with higher $f_{\rm out}$. The only exception is the CCCD results for NGC 4593, where the $\sigma_{\rm obs}$ for $f_{\rm out} = 0.1$ and $f_{\rm out} = 0.2$ are almost identical. The error estimate $\sigma_{\rm est}$ stays nearly the same for {\tt JAVELIN}, while it slightly increases at higher $f_{\rm out}$ for CCCD and CCPD. The ratio $\eta$ consistently drops when $f_{\rm out}$ increases, except for the CCPD results for NGC 4593. Both algorithms are more likely to underestimate the lag uncertainties for large number of outliers, although this brings the ratio $\eta$ for ICCF closer to unity from significantly overestimating the uncertainties in the baseline case. Like the previous cases, the ICCF results are less sensitive to the large outlier fraction, since the mistakes in single-epoch uncertainties affect {\tt JAVELIN} directly but affect ICCF only indirectly.

\subsubsection{Correlated Errors} \label{sssec:anl_corlerr}
\begin{figure*}
\includegraphics[width=\linewidth]{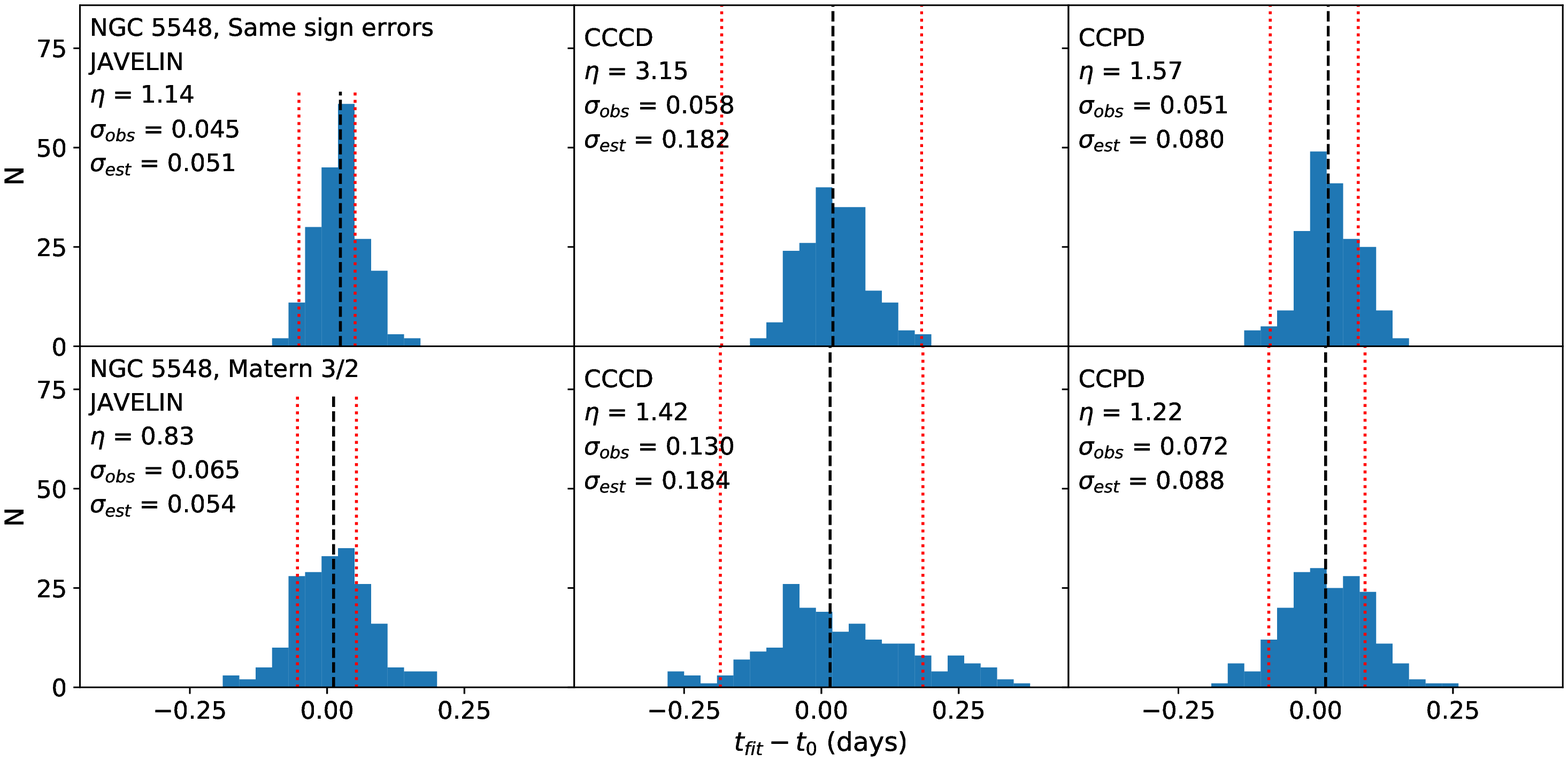}
\caption{Same as Figure \ref{fig:anl_ini} but for different configurations. The top row shows the results from making the errors of the continuum lightcurves and the line lightcurves the same sign. The bottom row is for the case where we model the correlated errors with the Matern 3/2 model.}
\label{fig:anl_errcorl}
\end{figure*}

Correlations between the single-epoch errors may also affect the lag uncertainty estimates. One approach to simulating correlated errors is to make the noise added to the corresponding epochs of the continuum lightcurves and the line lightcurves have the same sign. For example, if the noise added to an epoch of the continuum lightcurve is positive, then we require that the noise added to the line epoch for that date is also positive, although the amplitude can be different. In this case, the single-epoch errors between the continuum lightcurves and the line lightcurves are correlated. This effect can be created in RM campaigns by flux calibration errors which bias the lightcurve errors toward the same direction. The top row of Figure \ref{fig:anl_errcorl} shows the results with this error correlation. The parameters for both {\tt JAVELIN} and ICCF are generally consistent with the baseline configuration. The only exceptions are the CCCD results for NGC 4151 where the observed scatter $\sigma_{\rm obs}$ drops by about 25\% and the ratio $\eta$ increases by about 30\%. The overall behaviour of the algorithms with such correlated error indicates that it has little impact on the lag measurements. 

\begin{figure*}
\includegraphics[width=\linewidth]{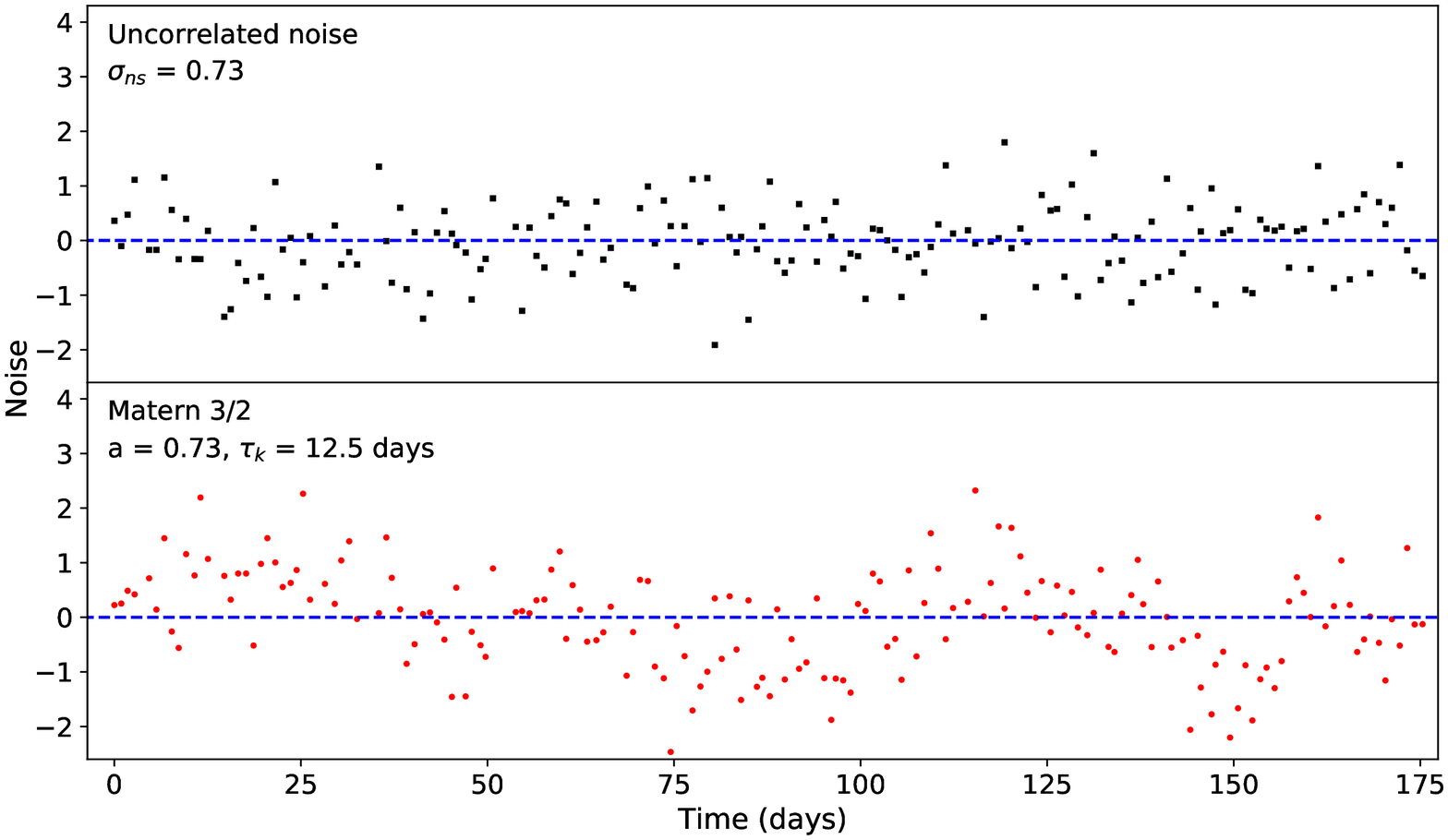}
\caption{Comparison of the noise added to the simulated continuum lightcurve for NGC 5548 based on uncorrelated Gaussian errors (upper panel) and correlated errors from the Matern 3/2 model (lower panel). The upper left corner of each panel shows the parameters of noise models (Equation \ref{eq:mtx_nsadd} and \ref{eq:ma32}). The blue dashed lines are drawn where the added noise equals zero.}
\label{fig:nsadd_errcorl}
\end{figure*}

Another approach to adding correlated errors is to generate the noise with a Gaussian process. The method of adding noise in the baseline configuration is equivalent to a Gaussian process specified by a covariance matrix $N_{ij} = \sigma_{\rm ns}^2 \, \delta_{ij}$, where $\sigma_{\rm ns} = X_i \langle \sigma_{i} \rangle$ and $\delta_{ij}$ is the Kronecker delta function. We can make the noise correlated by adding non-zero off-diagonal terms
\begin{equation}
N_{ij} = \sigma_{\rm ns}^2 \, \delta_{ij} + k(t_i,t_j) \, .
\label{eq:mtx_nsadd}
\end{equation}
Here we model the correlated errors using a 3/2 power Matern kernel \citep{Matern1960}
\begin{equation}
k(t_i,t_j) = a^2 \left( 1 + \frac{\sqrt{3}|t_i - t_j|}{\tau_{\rm k}} \right) {\rm exp}\left (-\frac{\sqrt{3}|t_i - t_j|}{\tau_{\rm k}} \right) .
\label{eq:ma32}
\end{equation}
Some studies on exoplanet transits \citep[e.g.,][]{Johnson2015} use this kernel to model the correlated errors in transit lightcurves due to the variability of the host star. The parameter $a$ characterizes the amplitude of the correlated errors and the parameter $\tau_{\rm k}$ describes the time scale on which the errors are correlated. Here we adopt $a = \sigma_{\rm ns}$ and $\tau_{\rm k} = 0.1\tau_{\mbox{\tiny DRW}}$. We separately add the correlated noise generated through this method to the continuum and line lightcurves but then assume the standard diagonal noise matrix for the algorithms. Figure \ref{fig:nsadd_errcorl} illustrates the difference between uncorrelated Gaussian noise and the correlated noise produced by the Matern 3/2 process. This error correlation model makes the noise tend to have the same sign on time scales of $\tau_{\rm k}$.

We show the results for this correlated noise model in the bottom row of Figure \ref{fig:anl_errcorl}. For both {\tt JAVELIN} and ICCF, there is no significant change in the median of $(t_{\rm fit}-t_0)$. The estimated uncertainties $\sigma_{\rm est}$ increase by about 30\% for the NGC 4151 CCPD results. Otherwise the estimated uncertainties $\sigma_{\rm est}$ are generally consistent with the baseline. The observed scatter $\sigma_{\rm obs}$ generally becomes significantly larger, and the ratio $\eta$ drops as a result. The change in $\sigma_{\rm obs}$ and $\eta$ is most significant for the CCCD method and for NGC 4151. Unmodeled correlated noise appears to broaden the $(t_{\rm fit}-t_0)$ distribution and cause a non-negligible drop in $\eta$ (i.e., it makes the algorithms more likely to underestimate the lag uncertainty). These temporally correlated errors have a bigger effect than the random outliers, because they are effectively a distortion in the lightcurve shapes. This means they can act like a violation of the assumptions of Equation (\ref{eq:linelc}) that the line lightcurve is a smoothed and delayed version of the continuum. We explore this further in Section \ref{subsec:anl_varbkg} where we explicitly add additional variability to the lightcurves.

\subsection{Effect of Transfer Functions} \label{subsec:anl_transtype}

In reality, the transfer function $\Psi(\tau)$ is not the top-hat function that we have assumed so far and is used in {\tt JAVELIN}. \citet{McHardy2018} obtained a transfer function consisting of a strong peak followed by a long tail between the X-ray, UV and optical bands in NGC 4593. \citet{Horne2019} recovered the emission-line transfer functions in NGC 5548, which generally show peaks at short lags and minor bumps at longer lags. Previous examinations of changing the transfer functions \citep[e.g.,][]{Rybicki1994,Zu2013} have found little effect on lag estimates.

\begin{figure}
\includegraphics[width=\linewidth]{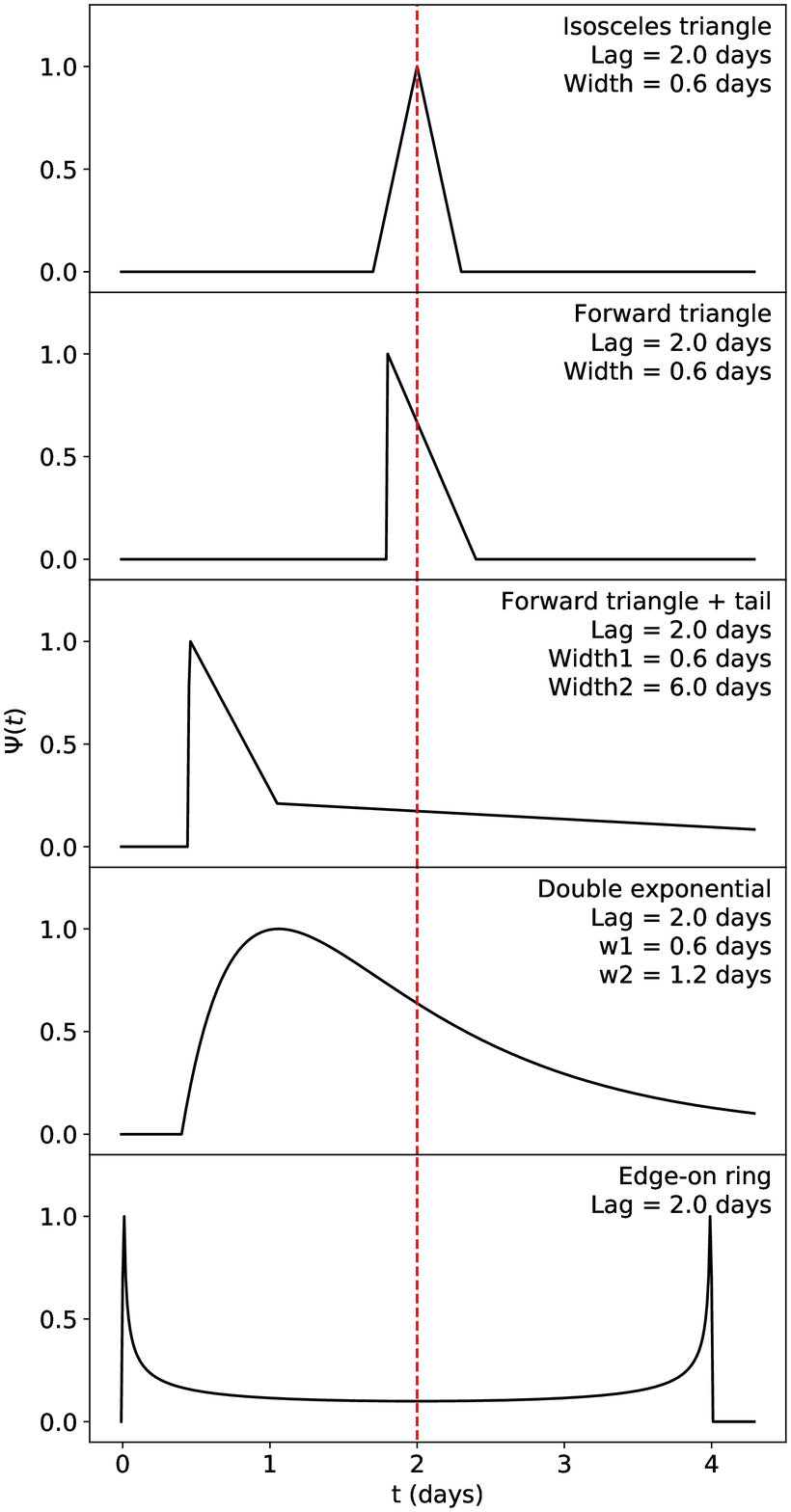}
\caption{Examples of the isosceles-triangular transfer function (top), the forward-triangular transfer function, the triangular transfer function with a long tail, the double-exponential transfer function and the transfer function of an edge-on ring (bottom). The transfer function parameters are given in each panel. The transfer functions are normalized to have the same peak value for visibility. The red dashed line is drawn at 2 days, the mean lag of the transfer functions.}
\label{fig:transfunc}
\end{figure}

Here we use five transfer functions other than the top hat, including an isosceles triangle, a ``forward'' triangle, a combination of two forward triangles to produce a narrow peak with a long tail, a combination of two exponentials and the transfer function of an edge-on ring. We set the width of the single triangular transfer functions to be the same as the width of the previous top-hat function and use the same mean lags as in the baseline configurations. For the combination of two triangles, we set the width of the second triangle to be 10 times the width of the top hat, so the function looks like a forward triangle followed by a long tail. The double-exponential transfer function has the analytic form
\begin{equation}
\Psi(t) =  A (1 - e^{-x}) e^{-y}\, ,
\label{eq:trans_dubexp}
\end{equation}
where $x = (t-t_0)/w_1$ and $y = (t-t_0)/w_2$. We adopt $w_1$ equal to the top-hat width and $w_2 = 1.2$ days so that the function has similar width to the double-triangular transfer function. We unify the function with $A = (w_1 + w_2)/w_2^2 \approx 1.26$, and the time offset $t_0$ is determined given $w1$, $w2$ and the mean lag $\langle \tau \rangle$. The normalized transfer function of an edge-on ring has the analytic form
\begin{equation}
\Psi(t) = \frac{1/\pi}{\sqrt{t(2\langle \tau \rangle - t)}}\, ,
\label{eq:trans_edgring}
\end{equation}
where $\langle \tau \rangle$ is the mean lag. Figure \ref{fig:transfunc} shows examples of the four transfer functions which by construction all have a mean lag of 2 days. In making these comparisons, it is important to use the correct mean lags (Equation \ref{eq:meanlag}) for the different transfer functions. While this is not crucial for the symmetric transfer functions, the mean lag for asymmetric transfer functions is not at the midpoint. 

\begin{figure*}
\includegraphics[width=\linewidth]{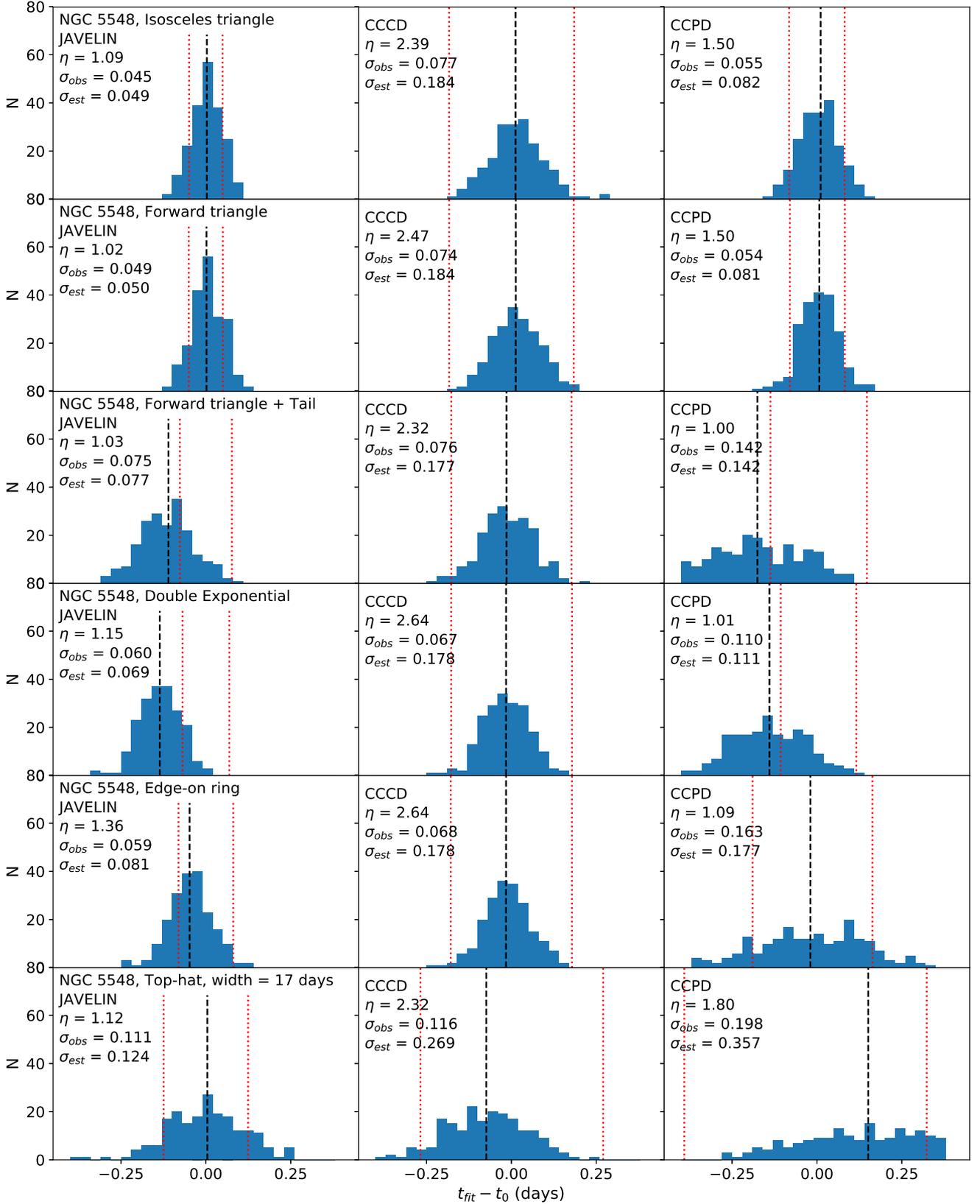}
\caption{Same as Figure \ref{fig:anl_ini} but for different transfer functions.}
\label{fig:anl_transtype}
\end{figure*}

We show the results in Figure \ref{fig:anl_transtype}. For the isosceles and forward triangles, the medians of the $(t_{\rm fit}-t_0)$ distribution remain close to zero. The scatter $\sigma_{\rm obs}$, the estimated uncertainty $\sigma_{\rm est}$ and the ratio $\eta$ are generally consistent with the baseline configurations. For the forward triangle with a long tail and the double-exponential transfer function, the algorithms tend to systematically underestimate the lag in the sense that the median $(t_{\rm fit}-t_0)$ is negative, although the systematic shifts are small relative to the input lags. This is not surprising because by more heavily smoothing the lightcurve, it is more difficult to detect the tail than the peak, which will tend to give the narrow peak more weight and lead to the bias. The observed scatter $\sigma_{\rm obs}$ and the estimated uncertainty $\sigma_{\rm est}$ increase in general, especially for {\tt JAVELIN} and CCPD, while $\eta$ stays nearly unchanged except for the NGC 4593 and Mrk 509 CCCD results and the NGC 5548 CCPD results. 

For the edge-on ring, the median $(t_{\rm fit}-t_0)$ of the NGC 4593 ICCF results deviates significantly from zero, while the others generally show similar behaviour to the baseline results. Both $\sigma_{\rm obs}$ and $\sigma_{\rm est}$ increase in most cases, especially for NGC 4593. The significantly larger $\sigma_{\rm obs}$ leads to small $\eta$ for the NGC 4593 ICCF results, indicating that ICCF does not work well in this specific case. This is not surprising. NGC 4593 has the shortest observational baseline and the large temporal width of the edge-on ring transfer function leads to a significant smoothing of the lightcurve variability. In general, the error ratio $\eta$ does not change significantly for the {\tt JAVELIN} and the other ICCF results. Overall, the form of the transfer function is not critical to the lag measurement for either algorithm.

The widths of the top-hat and the triangular transfer functions used above were small relative to the observational time baseline. We tried a top-hat transfer function with a width $w$ that is roughly 10\% of the temporal length of the lightcurve. We adopt $w = (17,7,3,27)$ days for NGC 5548, NGC 4151, NGC 4593 and Mrk 509, respectively. This leads to much more smoothing of the line lightcurve relative to the continuum. We use a random lag between $w/2$ and $w/2+2$ days for each realization. The bottom row of Figure \ref{fig:anl_transtype} shows the results for NGC 5548. The median $(t_{\rm fit}-t_0)$ shows larger deviation from zero relative to the baseline results for ICCF, while it does not change significantly for JAVELIN. The observed scatter $\sigma_{\rm obs}$ and the estimated uncertainty $\sigma_{\rm est}$ both increase, while the ratio $\eta$ stays nearly unchanged. We also tried the forward-triangular transfer functions with these larger widths. We got similar results except for the systematic shift due to the more weighted peak than the tail as discussed above. However, the shift is only $\sim 3$\% of the overall width of the transfer function. More strongly smoothing the lightcurve increases the uncertainties as expected, but the qualitative properties of the algorithms are unchanged. We expect this would hold if repeated for the other model transfer functions. 

\begin{figure*}
\includegraphics[width=\linewidth]{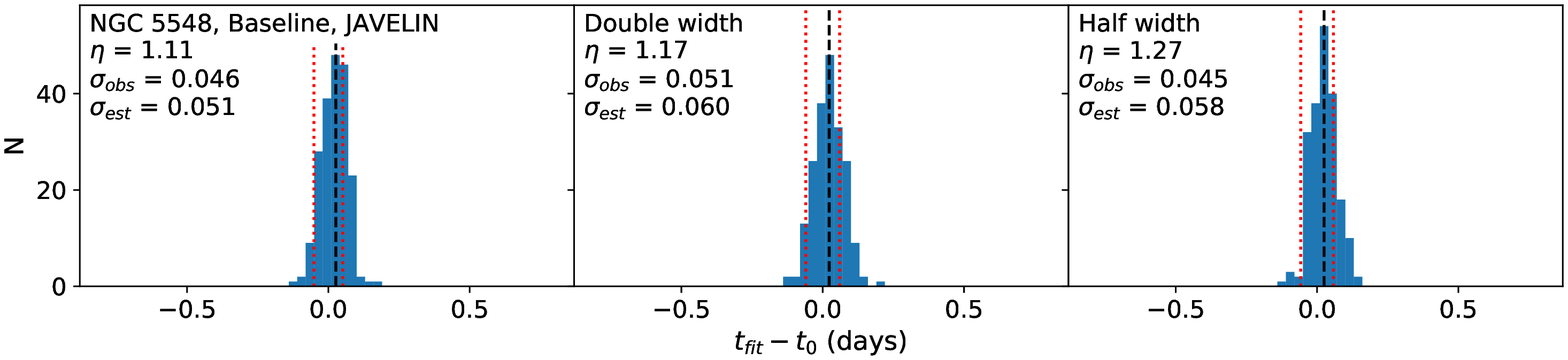}
\caption{Same as Figure \ref{fig:anl_ini} but for the {\tt JAVELIN} results with fixed top-hat width. The left panel shows the {\tt JAVELIN} results for the baseline configurations (same as the first panel of Figure \ref{fig:anl_ini}). The middle and right panel show the {\tt JAVELIN} results with the top-hat width fixed to twice and half the input values, respectively.}
\label{fig:anl_pfix_jav}
\end{figure*}

{\tt JAVELIN} assumes a top-hat transfer function and fits for the top-hat width and scale in addition to the lag. While {\tt JAVELIN} is generally able to recover the input lag, it usually cannot accurately recover the top-hat width. Essentially, the top-hat width and the scale factor between the line and continuum lightcurves are roughly degenerate when fitting typical data \citep[see][]{Zu2011}. In order to probe whether the large uncertainties in these parameters affect the lag measurements, we fit the simulated lightcurves in the baseline configurations with the top-hat width fixed to twice/half the input value. Figure \ref{fig:anl_pfix_jav} shows the results from NGC 5548. There is no significant change in the median $(t_{\rm fit}-t_0)$, the observed scatter $\sigma_{\rm obs}$, the estimated uncertainty $\sigma_{\rm est}$ and the ratio $\eta$. We also tried fixing the scale to incorrect values and obtained similar results.

\subsection{Effect of the Stochastic Process} \label{subsec:anl_simtype}

Several studies of Kepler lightcurves found that AGN variability deviates from the DRW model and can have a steeper PSD on time scales shorter than $\sim$ month \citep[e.g.,][]{Mushotzky2011,Kasliwal2015,Smith2018}. \citet{Zu2013} also saw weak evidence of this in OGLE lightcurves. We use two methods to explore the effects of the deviations from the DRW model, particularly at short time scales. 

\subsubsection{Kepler Covariance Model} \label{sssec:anl_ke}
\begin{figure}
\includegraphics[width=\columnwidth]{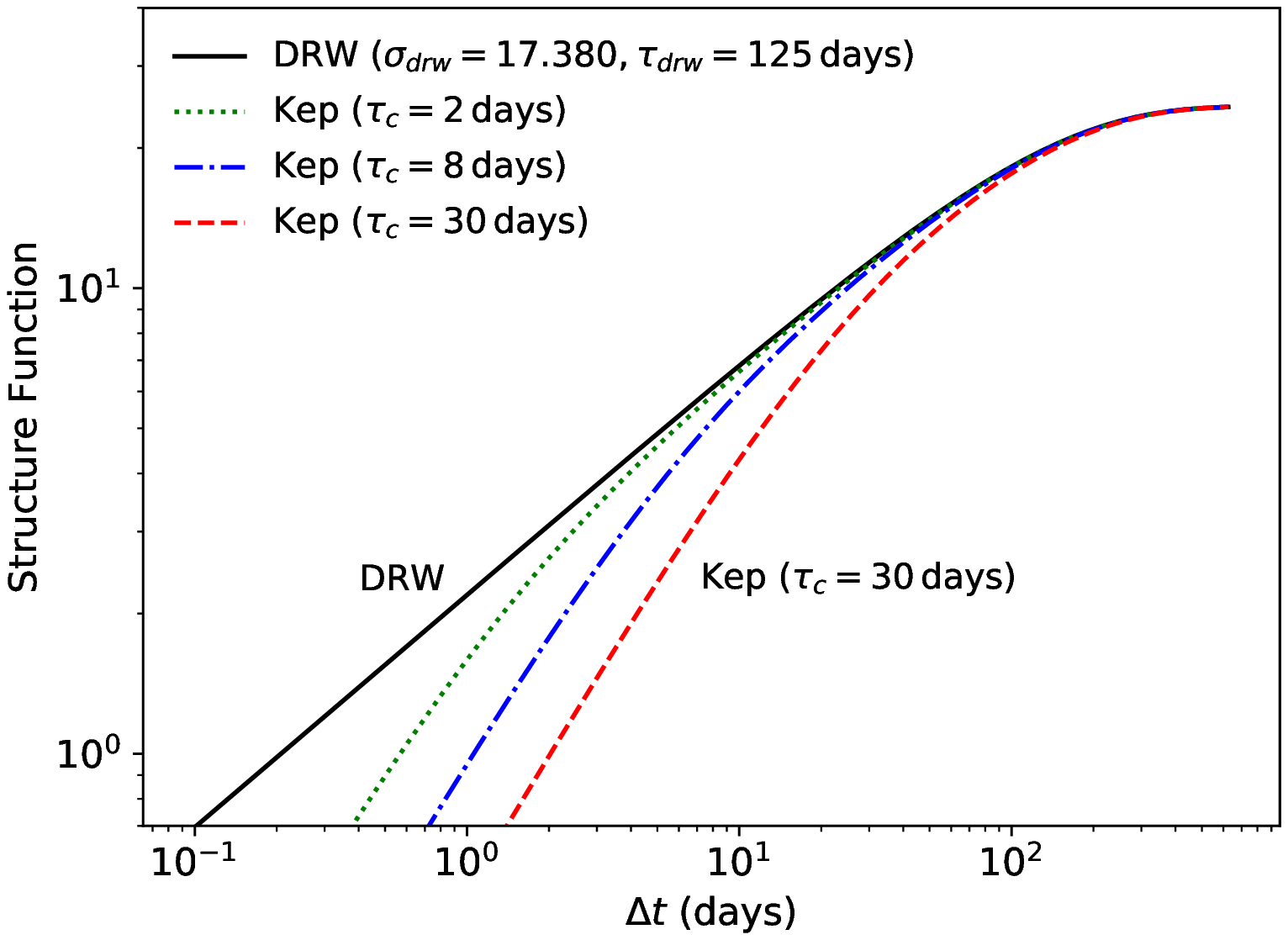}
\caption{Structure functions of the DRW and the ``Kepler'' covariance models for NGC 5548. The blue solid line represents the DRW, while the red dashed, blue dash-dotted and green dotted line represent the ``Kepler'' covariance model with $\tau_{\rm c} = 2,\, 8,\, 30$ days, respectively.}
\label{fig:sf_comp}
\end{figure}

In the first test we continue to use lightcurves constrained to resemble our four AGNs but generated using a different stochastic process. We use the ``Kepler'' process adopted by \citet{Yu2018}, with the covariance function 
\begin{equation}
S(\Delta t) = \sigma^2 \, [(1 + C) \, {\rm exp}(-|\Delta t / \tau_1|) - C \, {\rm exp}(-|\Delta t / \tau_2|)]
\label{eq:cov_ke}
\end{equation}
where $C = \tau_2 / (\tau_1 - \tau_2)$, $\sigma$ is an amplitude equivalent to $\sigma_{\mbox{\tiny DRW}}$ and $\tau_1$ is a time scale equivalent to $\tau_{\mbox{\tiny DRW}}$ in the DRW model. We can vary $\tau_2 < \tau_1$ to produce a cut-off in the structure function at short time scales. However, $\tau_2$ is not an intuitive indicator of the cut-off time scale, since the ``Kepler'' structure function starts to deviate from DRW at several times $\tau_2$. We therefore define a cut-off time scale $\tau_{\rm c}$ at which the ``Kepler'' structure function has 85\% the power of DRW. We adopt $\tau_{\rm c} = 2,\, 8,\, 30$ days and numerically solve for $\tau_2$ given each $\tau_{\rm c}$. Figure \ref{fig:sf_comp} compares the DRW and the ``Kepler'' structure functions. This covariance function allows a cut-off at a wider range of time scales than the ``Kepler-exponential'' model from \citet{Zu2013} without the problem of a non-positive definite matrix. 

\begin{figure*}
\includegraphics[width=\linewidth]{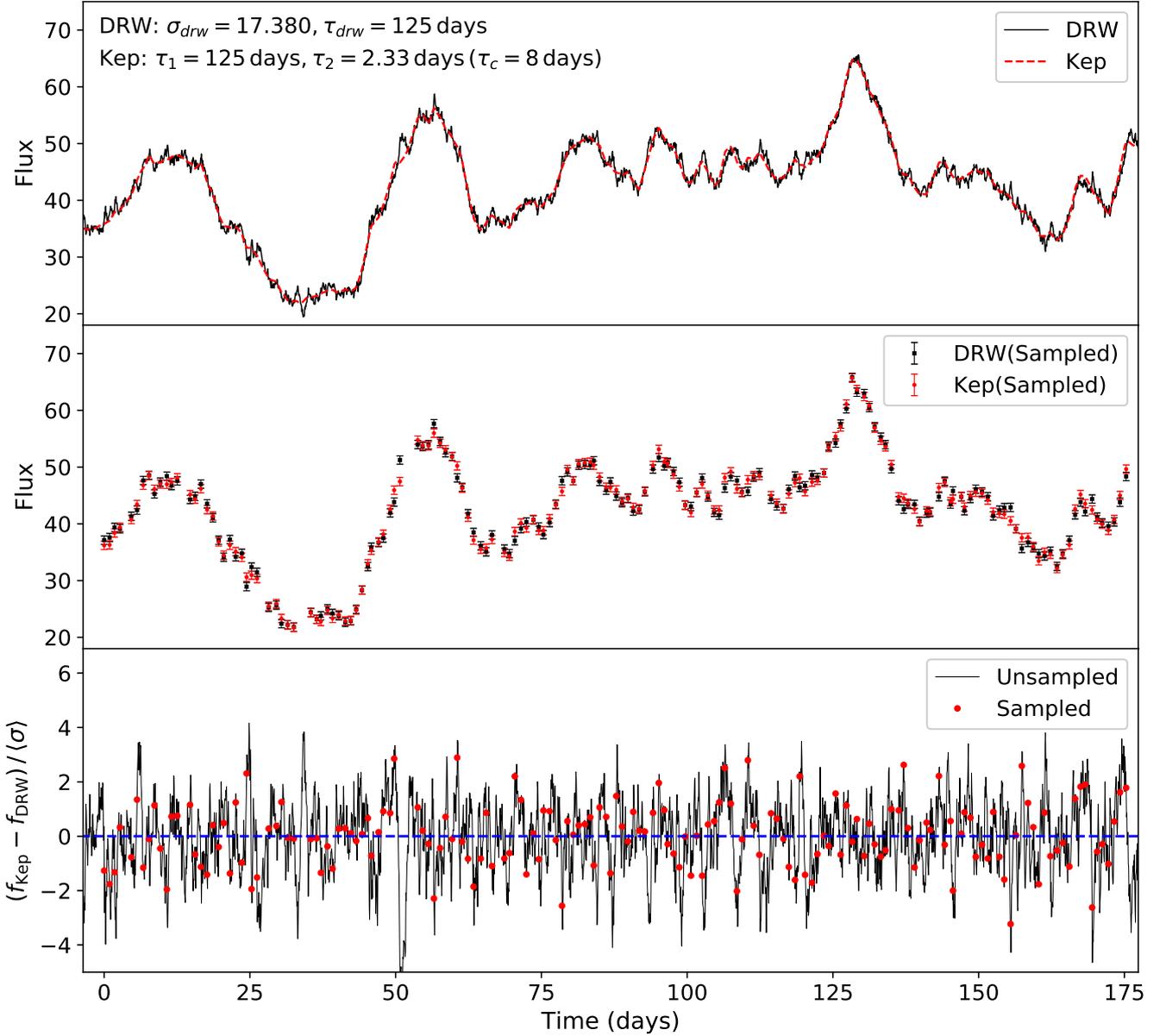}
\caption{Comparison of the simulated continuum lightcurve for NGC 5548 based on the DRW and the ``Kepler'' process models. The top panel shows the high-cadence noiseless lightcurves. The black solid line is the DRW lightcurve, while the red dashed line is the ``Kepler'' process lightcurve. The middle panel shows the lightcurves as they would be observed by the AGN STORM campaign in both cadence and noise. The black squares and red circles are the DRW and the ``Kepler'' process lightcurves, respectively. The bottom panel shows the residual from subtracting the DRW and the ``Kepler'' process lightcurves divided by the mean lightcurve errors. The black solid line and the red circles represent the residuals for the unsampled and sampled lightcurves, respectively.}
\label{fig:lc_kevsdrw}
\end{figure*}

We then create simulated lightcurves using the ``Kepler'' process with other parameters fixed to those in the baseline configuration. Figure \ref{fig:lc_kevsdrw} compares a realization of the DRW and the ``Kepler'' process lightcurves for NGC 5548 with $\tau_{\rm c} = 8$ days and using the same random seed so that the differences are only due to the change in the structure functions. The ``Kepler'' process lightcurve has less power at short time scales and is therefore smoother than the DRW lightcurve. However, after we resample and add noise to the lightcurves, the differences are rather subtle. 

\begin{figure*}
\includegraphics[width=\linewidth]{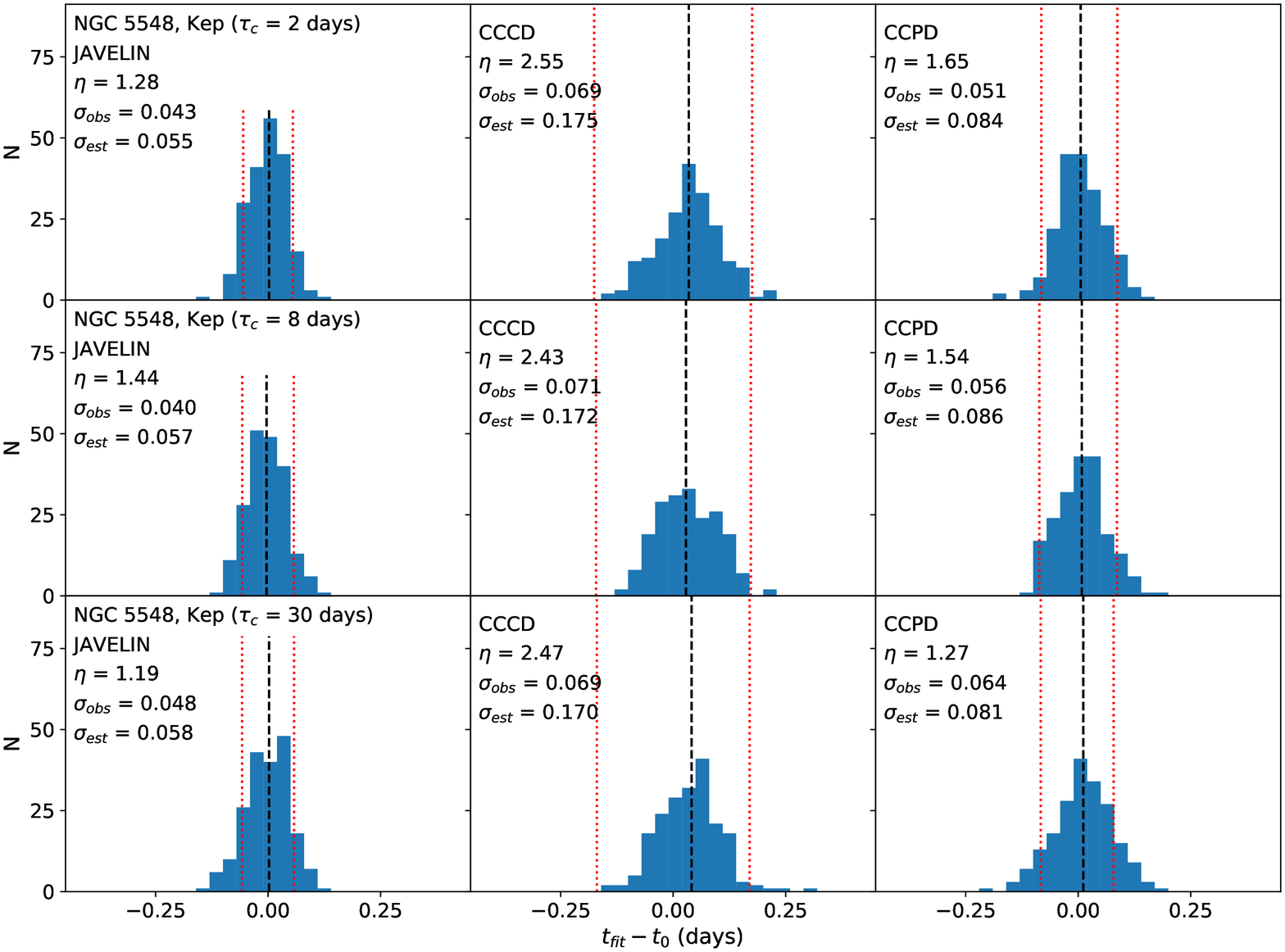}
\caption{Same as Figure \ref{fig:anl_ini} but for the ``Kepler'' covariance models.}
\label{fig:anl_radtype}
\end{figure*}

Figure \ref{fig:anl_radtype} shows the {\tt JAVELIN} and ICCF results for the ``Kepler'' process lightcurves. In most cases there is no strong variation in the median $(t_{\rm fit} - t_0)$, the observed scatter $\sigma_{\rm obs}$, the estimated uncertainty $\sigma_{\rm est}$ and the ratio $\eta$. The CCPD results for NGC 4151 give larger $\sigma_{\rm obs}$ and $\sigma_{\rm est}$ relative to the baseline, while the ratio $\eta$ stays nearly the same. For $\tau_{\rm c} = 8$ days, the ratio $\eta$ from {\tt JAVELIN} for NGC 5548 increases by about 30\% due to a slight drop of $\sigma_{\rm obs}$ and a slight rise of $\sigma_{\rm est}$. Overall, the deviations from the DRW model on short time scales do not have a significant impact on the lag measurements.

\subsubsection{Observed Kepler Lightcurve} \label{sssec:anl_kepobs}
\begin{figure*}
\includegraphics[width=\linewidth]{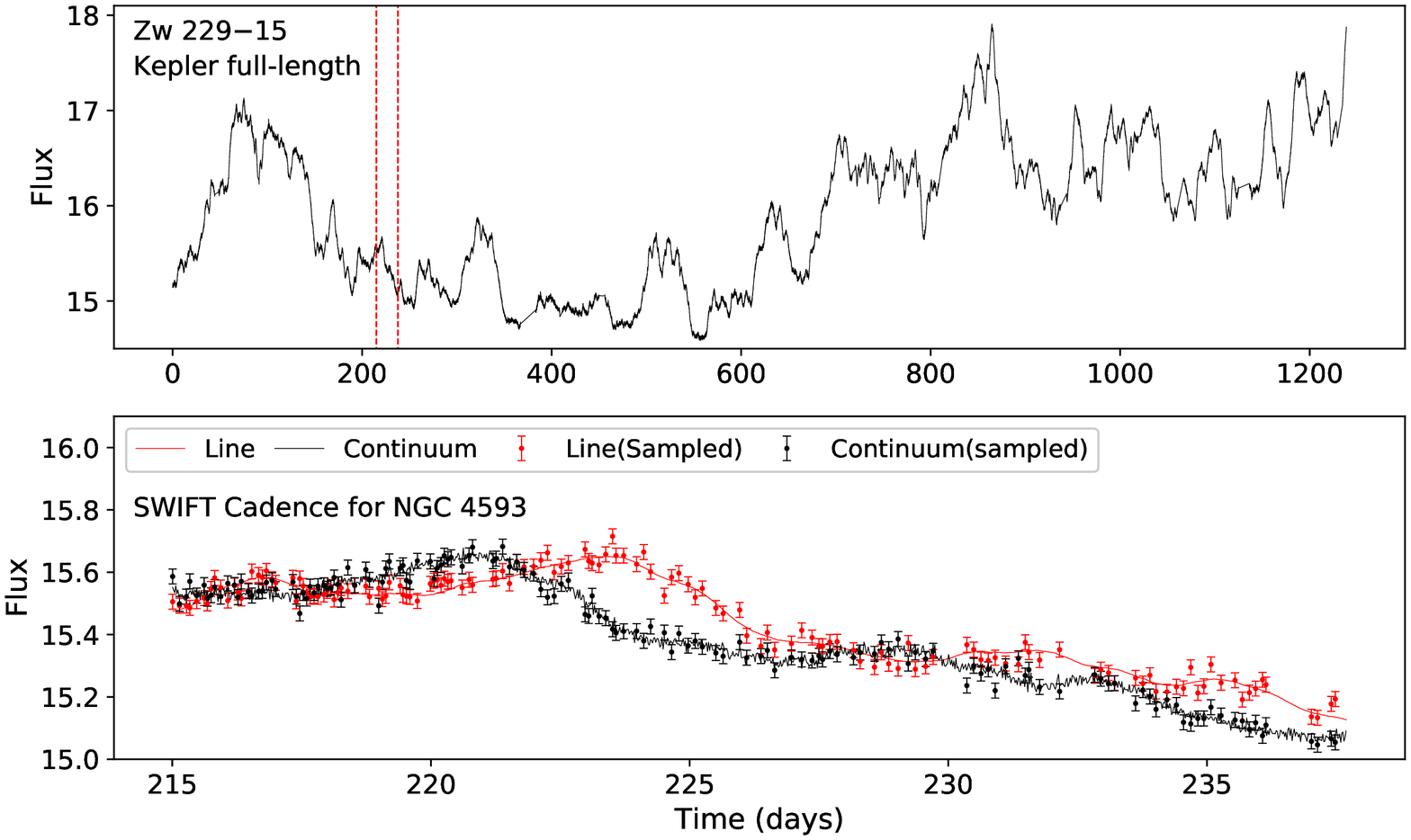}
\caption{The top panel shows the Kepler lightcurve of Zw 229$-$15 from \citet{Edelson2014}. We fill the observational gaps through linear interpolations. The bottom panel zooms in on the time interval between the red dashed lines in the top panel. The black solid line shows the observed Kepler lightcurve of Zw 229$-$15. The black points show the resampled Kepler lightcurve for the Swift cadence for NGC 4593 including noise. The red solid line shows an example of the simulated line lightcurves after convolving the Kepler lightcurve with a top-hat transfer function. The red points show the resampled line lightcurve using the fractional errors of the Swift data.}
\label{fig:lc_kepobs}
\end{figure*}

Our second test is to use the Kepler lightcurve of Zw 229$-$15 \citep{Edelson2014} shown in the top panel of Figure \ref{fig:lc_kepobs}. We select four time intervals within the Kepler baseline that have the same length as the Swift observations of NGC 4151 or NGC 4593, where few epochs within these intervals lie in the gaps of the Kepler lightcurve. We do not use the observations of NGC 5548 or Mrk 509 because their time baselines are too long to fit into a single Kepler quarter. In each time interval, we resample the Kepler lightcurve to the cadence of the Swift observations and use it as the simulated continuum lightcurve. We assign uncertainties to the resampled epochs so that the ratio of the single-epoch uncertainty to the standard deviation of the lightcurve is the same as the Swift lightcurves. The simulated continuum lightcurve in each interval can be viewed as an independent ``realization'' of the observed Kepler lightcurve. We then create 200 simulated line lightcurves for each of the four ``realizations'' following the procedures in Section \ref{sec:method}. The bottom panel of Figure \ref{fig:lc_kepobs} shows an example of the Kepler-based simulated lightcurves. The lightcurve shows weaker variations on short time scales than the DRW model.

\begin{figure*}
\includegraphics[width=\linewidth]{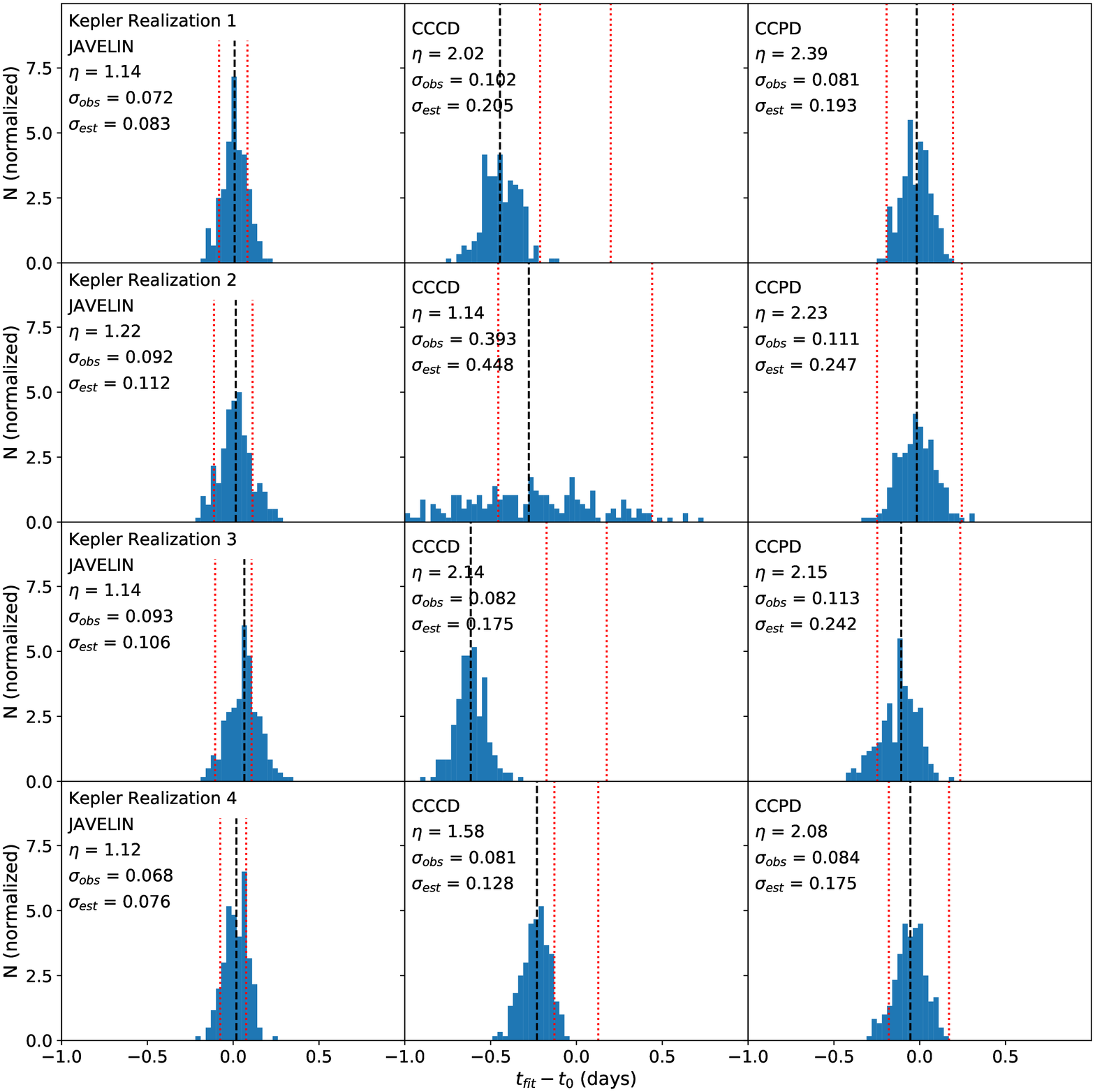}
\caption{Same as Figure \ref{fig:anl_ini} but for the simulated lightcurves based on the Kepler lightcurve of Zw 229$-$15. The first through the fourth row show the results from each ``realization'' of the Kepler lightcurves.}
\label{fig:anl_kepobs}
\end{figure*}

Table \ref{tab:results_kepobs} gives the {\tt JAVELIN} and ICCF results for the four ``realizations'' for these simulated lightcurves and Figure \ref{fig:anl_kepobs} shows the results for NGC 4593. The median $(t_{\rm fit} - t_0)$ generally stays close to zero, except the CCCD results for NGC 4593 and one realization of NGC 4151. It is not meaningful to directly compare the observed scatter $\sigma_{\rm obs}$ and the estimated uncertainties $\sigma_{\rm est}$ to the baseline results since the lightcurve shapes are different. However, the ratio $\eta$ still indicates the correctness of the lag uncertainty estimates. In most cases the ratio $\eta$ does not change significantly relative to the baseline results. This again indicates that any deviation of the continuum from the DRW assumption has little effect on the lag measurements. 

For some of the Kepler realizations, the CCCD results have an observed scatter $\sigma_{\rm obs}$ much larger than the other realizations. Most of these lightcurves show strong systematic trends, which can make it hard for the ICCF method to recover lags. If we detrend these lightcurves by fitting and subtracting a linear trend, CCCD generally shows better performance with smaller scatter $\sigma_{\rm obs}$ relative to the cases before detrending. The detrending also gives an $\eta$ ratio closer to the other realizations, and none of the realizations produce significantly different $\eta$ from the baseline results for the CCCD method after the detrending. We repeated this Kepler lightcurve test with additional tens of ``realizations'' for the NGC 4593 Swift cadence, and we got similar results except for the lightcurves where there is little variability after detrending and we do not expect a lag measurement. 

\subsection{Varying Backgrounds} \label{subsec:anl_varbkg}
RM makes the strong assumption that the line lightcurve is a smoothed and shifted version of the observed continuum lightcurve with a constant background level. However, \citet{Horne2019} found that this linear model fails for the observed lightcurves of NGC 5548, and they needed a time dependent background $L_0(t)$ instead of the constant background level $L_0$ in Equation (\ref{eq:linelc}) to obtain a good fit. This varying background may also explain the anomalous decoupling of the far UV continuum and the broad line variations found by \citet{Goad2016}. The most significant feature of the $L_0(t)$ found by \citet{Horne2019} is a drop from MJD 56740 to 56810 followed by a more rapid rise back until MJD 56840. The origin of this variation is not well understood. It may appear because the observed continuum is not the relevant extreme UV ionizing continuum, or due to the change of the line-of-sight covering factor of the obscurers absorbing the soft X-rays \citep[e.g.,][]{Mathur2017,Dehghanian2018,Goad2019,Kriss2019}.

We model this sort of behaviour by a set of Legendre polynomials. For a line lightcurve within time range $t_0<t< t_0 + t_{\rm m}$ (i.e., the original lightcurve spans 0 to $t_{\rm m}$ but we then add a lag of $t_0$), we model the background as
\begin{equation}
L_0(t) = \sum \limits_{i=1}^{N} a_i P_i(x) \, , {\rm where}\, x = 1 + 2(t - t_0 -t_{\rm m})/t_{\rm m} 
\label{eq:bkgvar}
\end{equation}
and $P_i(x)$ is the $i$th order Legendre polynomial. We adopt a maximum order $N=4$ and exclude the $0$th order so that $\langle L_0(t) \rangle = 0$. We choose the coefficients as 
\begin{equation}
a_i = r_i  \sigma_{\rm bkg} \sqrt{\frac{2i+1}{N}}
\label{eq:bkgvar_ai}
\end{equation}
where $i$ is the order of the Legendre polynomial, $r_i$ is a Gaussian random variable with zero mean and unit dispersion and $\sigma_{\rm bkg}$ is the desired standard deviation in $L_0(t)$. We choose this normalization so that each order contributes equally to $\sigma_{\rm bkg}$. We then linearly detrend $L_0(t)$ using the starting and ending point of the background lightcurve to avoid adding a strong systematic trend that can affect lag measurements even when also present in the continuum. The resultant standard deviation of $L_0(t)$ may differ from $\sigma_{\rm bkg}$ due to the random variable and the linear detrending, so we rescale $L_0(t)$ so that its standard deviation equals $\sigma_{\rm bkg}$. These choices lead to distortions that resemble those found for NGC 5548 in \citet{Horne2019}.

We considered two cases. In the first set of models, we generate two random backgrounds for each source and held the random seeds fixed. Here we expect to find a bias in the lag estimate. The observed scatter $\sigma_{\rm obs}$ can also increase, but the change would be less significant than the shift in the median $(t_{\rm fit}-t_0)$. In the second set of models, we randomly vary the backgrounds in each realization while holding the standard deviation $\sigma_{\rm bkg}$ fixed. This mimics repeated measurements of the same AGN, and here we expect the median of the $(t_{\rm fit}-t_0)$ distribution to be close to zero, but the dispersion $\sigma_{\rm obs}$ to be considerably larger due to the scatter in the individual estimates of the lag created by the varying backgrounds. 

\begin{figure*}
\includegraphics[width=\linewidth]{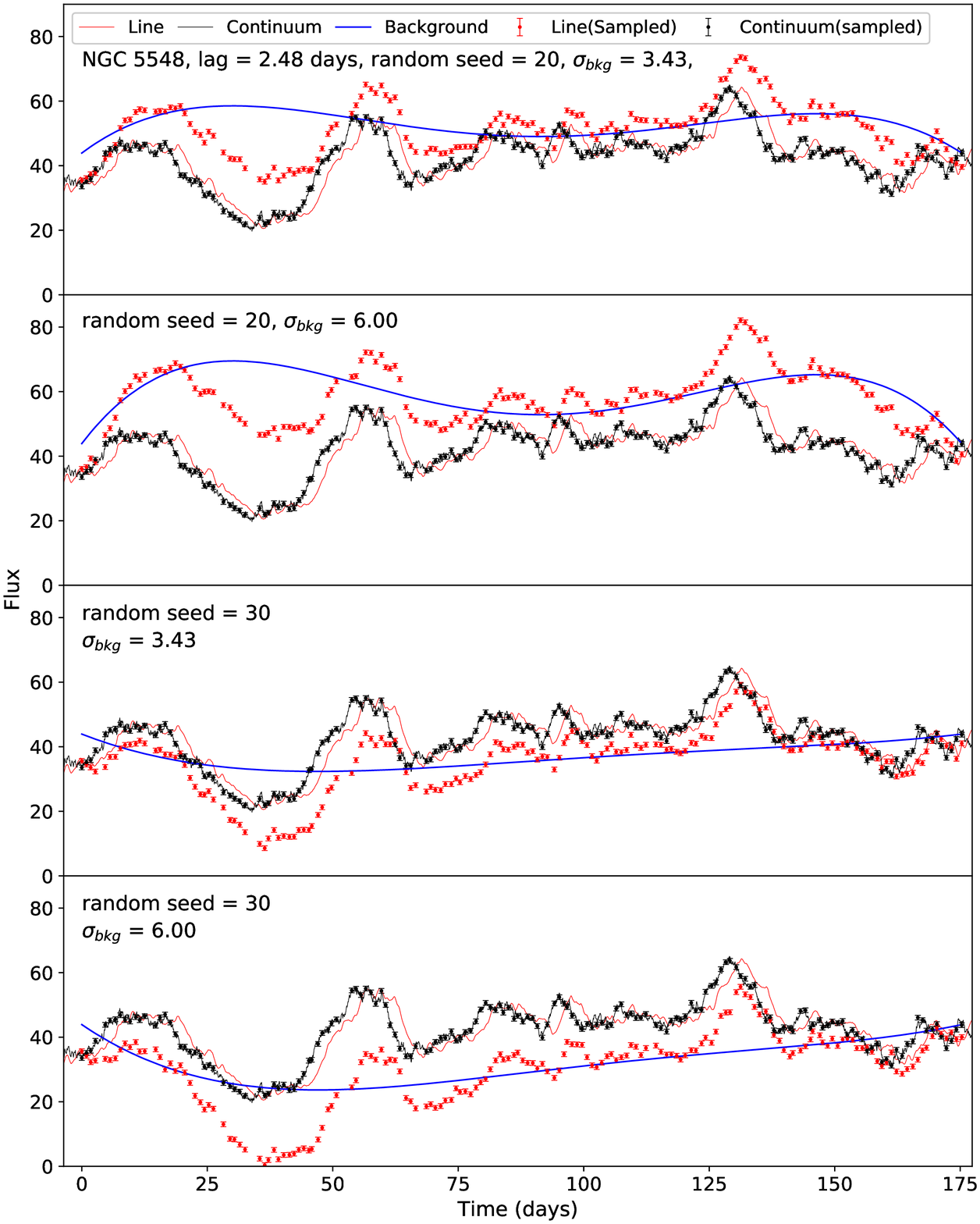}
\caption{Simulated continuum and line lightcurves for NGC 5548 with a background variation. Each panel shows one model configuration with the parameters in the top left corner. The flux is in arbitrary units. The solid black and red lines represent the high-cadence simulated lightcurve of the continuum and the emission lines, respectively, while the black and red points represent the resampled lightcurves. The blue solid line represents the background level. The background variation is only included in the resampled line lightcurve.}
\label{fig:lc_slbkg}
\end{figure*}

We first used two random seeds to generate the background lightcurve for each source. We set $\sigma_{\rm bkg}$ to 0.4 or 0.7 times the standard deviation of the observed lightcurve for each random seed. These ratios are typical of the background $L_0(t)$ used by \citet{Horne2019}. Figure \ref{fig:lc_slbkg} shows an example of the line lightcurve after adding a varying background. The lags of these lightcurves are likely to deviate significantly from the input due to the deviation of the resampled line lightcurves (red points) from the high-cadence lightcurve (red solid line) with a constant background. We therefore consider lags outside of the $-$2 days to 8 days range for the analysis here. 

\begin{figure*}
\includegraphics[width=\linewidth]{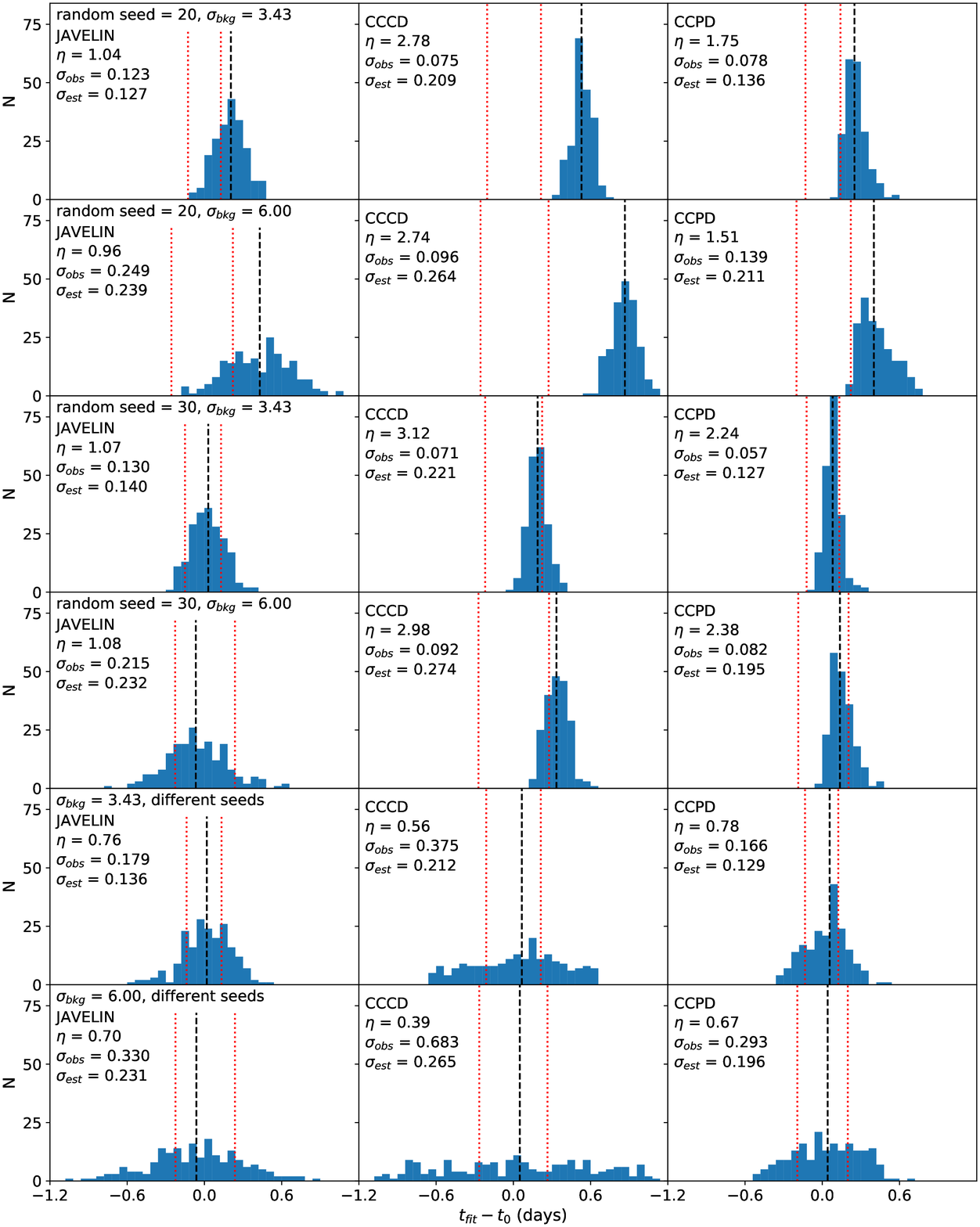}
\caption{Same as Figure \ref{fig:anl_ini} but for the results from background variations. Each panel shows the results from one model configuration with the parameters at the top left corner.}
\label{fig:anl_bkgvar}
\end{figure*}

Table \ref{tab:results_bkg} includes the model parameters and the {\tt JAVELIN} and ICCF results after adding the varying backgrounds. Figure \ref{fig:anl_bkgvar} shows the results for NGC 5548. In most cases the median of $(t_{\rm fit}-t_0)$ deviates significantly from zero. These shifts are also ``visible'' in the lightcurves. When the line lightcurve is rising, pulling the lightcurve down seems to move the lightcurve further right and leads to a larger lag. On the other hand, when the line lightcurve is declining, a drop in the lightcurve seems to move the lightcurve left and makes the lag smaller. The resultant median of $(t_{\rm fit}-t_0)$ is a balance between these two features. The scatter $\sigma_{\rm obs}$ and the estimated uncertainty $\sigma_{\rm est}$ increase significantly relative to the baseline. The ratio $\eta$ stays nearly the same for the NGC 5548 {\tt JAVELIN} results, but otherwise does not show a consistent pattern. Both algorithms are likely to give incorrect lags and uncertainties after adding the background variation. {\tt JAVELIN} is generally more sensitive to this for the lag uncertainties $\sigma_{\rm obs}$ and $\sigma_{\rm est}$, while ICCF, especially CCCD, is more sensitive to this for the median $(t_{\rm fit}-t_0)$.

In the second set of models, we randomly change the backgrounds in each trial while holding the standard deviation $\sigma_{\rm bkg}$ fixed. We show the results in the bottom two rows of Figure \ref{fig:anl_bkgvar}. As expected, the medians of $(t_{\rm fit} - t_0)$ are generally closer to zero than in the fixed random seed cases. The observed scatter $\sigma_{\rm obs}$ increases significantly, while most of the estimated uncertainties $\sigma_{\rm est}$ changes only slightly. The ratio $\eta$ drops as a result, and both algorithms underestimate the lag uncertainties except in a few cases for NGC 4593. As noted earlier, the linear parameters $L\tbf{q}$ in {\tt JAVELIN} allows for the modeling of systematic trends. The trends can be different for the continuum and line lightcurves, allowing {\tt JAVELIN} to handle the problem considered in this section as part of its analysis. However, an expansion of our analysis to fully explore these modifications is beyond our present scope.

\section{Discussion and Summary} \label{sec:summary}
We used observationally-constrained simulated lightcurves to probe the effects of systematic errors on the {\tt JAVELIN} and ICCF methods under a wide range of circumstances. We measured the lags from the simulated lightcurves through {\tt JAVELIN} and ICCF and compared the input lag $t_0$ and the output lags $t_{\rm fit}$. We characterized the performance of the algorithms with the median $(t_{\rm fit}-t_0)$, the observed scatter $\sigma_{\rm obs}$, the estimated uncertainty $\sigma_{\rm est}$ and the ratio $\eta = \sigma_{\rm est} / \sigma_{\rm obs}$.

In general we found that both methods are reasonably robust to the presence of all but one of the systematic problems we explored. In most circumstances, {\tt JAVELIN} produces better lag error estimates in the sense that its error estimates are more consistent with the scatter of the results from random trials (i.e., the ratio $\eta$ closer to unity). The ICCF method tends to overestimate the lag uncertainties. Because the ICCF method overestimates uncertainties when there are no severe systematic problems, it can be somewhat more ``robust'' when there are severe systematic problems. 

Incorrect single-epoch error estimates and correlated errors in the lightcurves can lead to incorrect lag uncertainties, but generally not by large factors unless there are very big problems. Because {\tt JAVELIN} is explicitly Gaussian, its error estimates are directly affected by problems in the lightcurve uncertainty estimates. If the true uncertainties are twice or half the uncertainties supplied to {\tt JAVELIN}, it will get a lag uncertainty wrong by a factor of two simply because of its mathematical structure. Since the ICCF method does not explicitly depend on the single-epoch errors, the effects of the problems in the lightcurve errors tend to be more subtle. Temporally correlated errors can have a bigger effect than the random errors, probably because they are effectively a distortion in the lightcurve shape. 

As previously found by \citet{Rybicki1994} and \citet{Zu2013}, changes in the shape of the transfer function have little effect on the lags. The primary exception is that a transfer function with a narrow peak and long tails will increasingly favor the lag due to the peak as the tail becomes longer. However, this effect was modest even for the 10:1 time scale ratio we considered in our experiments.

As we would expect from the underlying mathematics of {\tt JAVELIN}, it does not matter if the true stochastic process of the continuum lightcurves differs from the DRW model used by {\tt JAVELIN}. We demonstrate this both with model lightcurves that have suppressed power on short time scales and with empirical lightcurves from Kepler which show such modified structure functions. The performance of the ICCF method also shows no significant consequences from changes in the process driving the variability. 

As noted in the introduction, there are also many studies exploring how the algorithms perform as the cadence, temporal baseline and signal-to-noise ratio of the observations change, and address the likelihood of lag measurements for lower-cadence lightcurves \citep[e.g.,][]{Horne2004,King2015,Shen2015,Yu2018,Li2019}. The more recent studies generally find that {\tt JAVELIN} is more likely to yield an accurate lag measurement and, consistent with our results, that it generally provides more accurate lag uncertainty estimates. In general, however, these studies have generated their simulated lightcurves using {\tt JAVELIN}'s baseline assumptions, which is why we have focused on the consequences of violating those assumptions. 

We do, however, identify one systematic problem which produces significant biases. The standard assumption of RM is that the line lightcurve is a smoothed and shifted version of the continuum (Equation \ref{eq:linelc}). If this assumption is incorrect, then both {\tt JAVELIN} and ICCF begin to produce increasingly inaccurate lag estimates. We observe such effects after adding extra variability to the simulated lightcurves that resembles the anomalous variability found in NGC 5548. Such violations of the fundamental assumptions of RM are probably the dominant cause of problems in lag estimates from lightcurves which show variability features that should otherwise yield accurate lag measurements. We did not test combinations of multiple systematic errors because of the combinatoric explosion of cases. Mathematically there should be no surprises and the varying background effect will remain the most important source of systematic errors for both algorithms. While we discuss our results mostly in terms of the emission line RM, they are equally applicable to continuum RM. The tests we performed for JAVELIN and ICCF can also be extended to other algorithms such as ZDCF and CREAM, or to the measurement of time delays in gravitational lenses \citep[e.g.,][]{Liao2015} for prospective future studies.

\section*{Acknowledgements}
%

We thank the anonymous referee for the comments and suggestions on our paper. C.~S.~Kochanek is supported by NSF grants AST-1515876, AST-1515927 and AST-1814440. Z.~Yu is supported by NSF under Grant No. 1615553. E.~M.~Cackett gratefully acknowledges support from NSF grant AST-1909199. I.~M.~McHardy acknowledges support from STFC under grant ST/R000638/1.




\bibliographystyle{mnras}
\bibliography{ms} 








\bsp	
\label{lastpage}
\end{document}